\documentclass[11pt]{article}
\usepackage[colorlinks,linkcolor=blue,citecolor=blue,urlcolor=blue,bookmarks,bookmarksnumbered]{hyperref}
\usepackage{amsmath, amssymb}
\usepackage{cite}

\numberwithin{equation}{section}

\renewcommand{\leq}{\leqslant}

\DeclareMathOperator{\Tr}{\textrm{Tr}}
\newcommand{\K}{\mathbb{K}}
\newcommand{\IP}{\mathbb{P}}

\newcommand{\X}{\mathbb{X}}
\newcommand{\rd}{\mathrm{d}}
\newcommand{\rD}{\mathrm{D}}
\newcommand{\GN}{\mathord{G_{\mkern-4muN\mkern-2mu}}}

\newcommand{\bra}[1]{\langle #1 |}
\newcommand{\ket}[1]{| #1 \rangle}
\newcommand\be{\begin{equation}}
\newcommand\ee{\end{equation}}
\newcommand\nn{\nonumber}
\newcommand\pa{\partial}

   \parskip\medskipamount     \lineskip=0pt
\thicklines \footnotesep=2mm
\begin{document}

\setcounter{page}{0}
\thispagestyle{empty}

\vspace*{3cm}

\begin{center}
\renewcommand{\thefootnote}{\fnsymbol{footnote}} 

{\Large\bf Quantum Spacetime, Quantum Gravity\\[1mm]
and Gravitized Quantum Theory%
\footnote{This talk, presented by Djordje Minic at the International Conference
{\em\/Nonlinearity, Nonlocality and Ultrametricity,} May 26--30, 2025, Belgrade, Serbia, 
is based on 
Refs.%
\cite{Minic:2020oho, Berglund:2022qcc, Berglund:2022skk, Berglund:2023vrm, Freidel:2022ryr, Berglund:2022qsb, Berglund:2023gur, Hubsch:2024agh, Berglund:2025qyo, Hur:2025lqc}.
}}

\vspace{5mm}

{\bf Tristan H\"ubsch}$\,^*$ and {\bf Djordje Minic}$\,^\dag$\\*[2mm]
{$^*$\slshape Department of Physics that Astronomy, Howard University\\ Washington, DC, 20059, USA; \texttt{thubsch@howard.edu}}\\
{$^\dag$\slshape Department of Physics, Virginia Tech,\\ Blacksburg, VA 24061, USA; \texttt{dminic@vt.edu}}

\vspace{15mm}

\parbox{12cm}{\emph{Abstract}.
 General relativity is a background-independent theory of a dynamical classical spacetime geometry. Quantum theory is formulated in a classical spacetime, as an intrinsically probabilistic, contextual theory of non-classical, interfering probabilities, with a fixed Born rule for computing those probabilities. We argue that the quantum nature of spacetime, which 
 includes a non-commutative dual companion to the (observed) classical
 spacetime, is the reason behind an intrinsically probabilistic and contextual nature of quantum theory, with the fixed Born rule.  
 In quantum gravity, we claim, quantum theory is gravitized into a background-independent structure with dynamical and contextual quantum probabilities. 
 This proposal implies intrinsic triple and higher-order interference in the context of massive quantum probes, 
 which
 sheds light on string theory and the observed vacuum energy as well as the masses of elementary particles.


}

\newpage
\begin{quote}
\baselineskip=12pt\parskip=0pt
\tableofcontents
\end{quote}

\setcounter{footnote}{0}
\end{center}

\section{Introduction}
\subsection{From Foundations to a Gravitized Quantum Spacetime}
{{2025 was the International Year of Quantum Science and Technology and the 100th birthday of quantum mechanics---100 years since Heisenberg's first paper on quantum mechanics from 1925.}}
It has been a remarkable century for quantum mechanics. Quantum theory precisely realizes the {{ancient atomic hypothesis}} and it is true for everything we know about matter: in biology, chemistry,  
condensed matter,
molecular, atomic, nuclear and elementary
particle physics. Some of the current frontiers of quantum theory include quantum biology, quantum gravity, and quantum cosmology.
Quantum physics is also the basis of major technological advances, without which our contemporary world would be unimaginable, such as semiconductors, modern electronics, lasers, atomic clocks, solar cells, medical imaging. The frontiers of practical uses of quantum theory are also many: quantum information and computation, quantum cryptography, quantum sensors, as well as others.

{{Yet, all these inherently 
quantum phenomena  
are understood as existing and unfolding
in classical spacetime. Is this (apparent) dichotomy between the quantum nature of matter and the classical nature of spacetime fundamental?}} Or, is it indicative of a major missing piece in the foundations of physics?
{{If indeed there is something major missing in our understanding of Nature, what new physics would capture such quantum (``atomic'') structure of spacetime, if that concept makes any empirical sense? (The idea of ``ameres" of ancient atomists, which can be viewed as ``atoms of space and time,'' comes close to the idea of quantum spacetime, at least, ``philosophically"---see~\cite{Hubsch:2026The}.)}} Is quantum theory as it is currently known the deepest description of physical reality? Is there ``life'' (that is, consistent physics) beyond quantum theory? And, what empirical phenomena would such ``superquantum'' (or ``metaquantum'') physics explain that are beyond the reach of current quantum theory 
and so warrant its introduction---%
just as quantum physics is necessary
to explain atomic phenomena, such as radioactive decay of atoms, or stability of atoms, that are simply beyond the realm of classical physics?

{\bf Motivation}: It is safe so say that quantum theory is the reigning theory of matter in classical spacetime. Currently, there exists \underline{\em no\/} conflict between quantum theory (including quantum field theory) and experiment. Similarly, general relativity is the reigning theory of classical spacetime with, in principle quantum, matter sources.
And currently, there is \underline{\em no\/} conflict between general relativity and experiment.

We note that spacetime physics is classical and background-independent, with fixed spacetime polarization and dynamical geometry of spacetime, whereas quantum physics is intrinsically probabilistic,
with a dynamical polarization and fixed geometry of the space of states, and the fixed rule for computing probabilities---the Born rule.
At the moment, quantum gravity (the consistent union of quantum theory and spacetime physics) \underline{\em does not\/} have any experimental basis (except, maybe,  
with regard to
vacuum energy and the cosmological constant problem)! 
Nevertheless,
searches for experimental 
evidence and verification
of quantum gravity are underway and some will be discussed in this paper,  
humbly presented in celebration of the 80th birthday 
of Prof.~Branko Dragovich.

{\bf Summary of the main points of the paper}: 
Spacetime (denoted $x$) physics is classical, non-probabilistic, but background independent (generally covariant). Quantum physics is fundamentally probabilistic, but with fixed/rigid Born rule. We will argue that {quantum gravity (QG) is in fact ``gravitized quantum theory'' (GQ) of quantum spacetime},
{with a dynamical and contextual Born rule, and an intrinsic triple and higher-order quantum interference---which provides a coveted and crucial experimental signature.} 
There are consequences of this new 
view of quantum gravity:
It will enable us to give formulae for the cosmological constant, Higgs mass, and masses and mixing matrices of quarks and leptons.
{We will also state our predictions for the neutrino masses}.
There also exists an {\em explicit realization\/} of QG=GQ: {the so-called {\em metastring theory\/}: 
an intrinsically non-commutative and T-duality covariant chiral and phase-space-like formulation of string theory (and its zero modes---{\em metaparticles\/})}, existing in and defining a {quantum (or, modular) spacetime}.
Dual particles and fields in this picture model (fuzzy) dark matter, and
dark energy is identified with the curvatures of the dual spacetime. {In addition, integrating out the directly unobservable dual spacetime (denoted $\tilde{x}$, with $[x,\tilde{x}]=i \lambda^2$), provides a new view and way to address the measurement problem and the relation between quantum and classical physics in classical spacetime, $x$.}

We will use the cosmological constant (cc) problem to argue for our ``quantum gravity = gravitized quantum theory'' ({QG=QG}) motto, 
and {emphasize the need for empirical probes} such as
{triple and higher-order quantum interference}, etc., which are at low energies essentially captured by the scale of vacuum energy.  
The dark energy sector emerges, we claim, from the curvatures of the dual spacetime, which is also 
at the heart of
the measurement problem (and its foreseeable resolution) 
and the foundations of quantum theory. Thus the measured scale of dark energy is, we propose, the scale at which gravitized quantum theory should be observed with massive quantum probes and such experiments are within reach! 
{We will emphasize that quantum theory stems from the compatibility of fundamental length/time and Lorentz symmetry. Thus, the intrinsic quantum-probabilistic nature of quantum phenomena stems from the quantum spacetime and its 
foundational agreement
with the conserved quantities, such as energy and momentum, as implied by the continuous symmetries of spacetime. The tension between the quantum and continuous nature of spacetime is resolved by interference of observer dependent spacetimes, and the intrinsically probabilistic description, based on a maximally symmetric probability measure---the rigid, quadratic Born rule, compatible with a unitary evolution of quantum states. 
In this description, the familiar and directly observable canonical spacetime ($x$) has a dual/conjugate partner%
---the dual spacetime ($\tilde{x}$), such that $[x, \tilde{x} ] = i \lambda^2$, 
where $\lambda$ denotes the fundamental length/time.
Inclusion of gravity then ``gravitizes'' the quantum theory, rendering the Born rule dynamical and contextual. This, apart from the quantum nature spacetime, is the key property of quantum gravity. 
This passage from quantum theory to the {\em gravitized quantum theory\/} is very similar to how special theory of relativity becomes ``gravitized'' into a general theory of relativity.
The problems of quantum measurement and the quantum-to-classical transition are captured by averaging over the dual spacetime coordinates. 

Fundamentally, quantum theory teaches us about:
({\bf1})~the quantum nature of spacetime and 
({\bf2})~a new kind of probability theory, which can be 
{\em a})~extended in the context of a dynamical spacetime, and 
{\em b})~which can also be used in other problems that go beyond the current areas of application of quantum physics (as well as the proposed new view on the problem of quantum gravity)---for example, in the realm of complex and adaptive systems.}

{\bf Quantum Theory and the Structure of Physics:}
Here we wish to say a few words about the structure of physics, and the central role of quantum theory and quantum gravity in it. Quantum theory (together with relativity) is crucial for the existing (albeit, incomplete) structure of {{simple, or fundamental, and elementary (as opposed to complex)}} physics,
characterized by the 3 fundamental constants $c$, $\GN$, $\hbar$ (captured by Planck's fundamental length {{$l_P=(\GN \hbar/c^3)^{1/2}$}}, time {{$t_P =l_P/c$}} and energy $E_P = \hbar/t_P$):

$\blacktriangleright\,c$ is {{speed of light}}: the fundamental conversion factor between space and time; or in our view ``the measure of how fast quantum information is transferred in quantum spacetime.''

$\blacktriangleright\,\GN$ is {{gravitational constant}}: it captures ``elasticity" of spacetime, measures how geometry reacts to matter; or in our view ``the measure of the curvature of general geometry of quantum probability.''

$\blacktriangleright\,\hbar$ is {{Planck's quantum constant}}: it measures ``virtuality'' of histories of a physical system; or in our view ``the conversion factor between general quantum probability and time at Planck energy.''

{{There are 6  
well-understood
combinations of the fundamental constants:
the origin (the $c\to0$, $\GN\to0$, $\hbar\to0$ limit),
{{$c$, $\GN$, $\GN c$, $\hbar$, $\hbar c$}}
(respectively: classical spacetime, general relativistic, with its special relativistic as well as Newtonian and Galilean limits, and quantum mechanics and quantum field theory of quantum matter, non-relativistic and relativistic); there are also two less-understood combinations: $\hbar\GN$, $c \GN \hbar$ 
(quantum spacetime, quantum matter)}}.
These represent, respectively, non-relativistic and relativistic theories of quantum gravity and we will comment on both in what follows.

The non-quantum ($\hbar \to 0$ limit) is represented by spacetime physics.
The story of spacetime physics is the story of classical space and time, as well as the power of symmetrization of the fundamental concepts of space and time.
Following Penrose, we may distinguish the following views on space and time:

{\bf{Aristotelian space and time}}: space and time separate (there are preferred frames due to the fundamental absence of relativity).

{\bf{Galilean space and time}}: space mixes with time, but time not with space (and so there is absolute time); in this case we encounter inertial frames and relativity.

{\bf{Newtonian space and time}}: curved (but locally Galilean) space with absolute time---gravity emerges as geometry (from a modern point of view).

{\bf{Minkowski's spacetime}}: space and time symmetrized (space mixes with time and time with space; observer dependent space and time as different ``slices" of spacetime); the Minkowski geometry of spacetime is fixed, $\rd s^2_M = \eta_{ab}\,\rd x^a \rd x^b$.

{\bf{Einstein's spacetime}}: curved spacetime, dynamical causality (spacetime symmetrization of Newtonian picture); gravity as geometry of spacetime. Therefore, we work with dynamical metric of general relativity (GR), $\rd s^2_{GR} = g_{ab}(x)\,\rd x^a \rd x^b$.

{{We emphasize that from the modern point of view, Newtonian gravitational physics 
can be understood from  
the geometric Newton--Cartan formulation, which is, essentially Einstein's theory in classical spacetime with an absolute time. Thus, the deeper theory (Einstein's) sheds light on its limits (Newton's), both
conceptually and practically. The correspondence principle is in operation when thinking about the Newtonian limit of Einstein's theory of gravity.}}

Quantum theory can be also understood as a story of symmetrization.
{This is represented by the {Koopman--von Neumann (KN) viewpoint on quantum and classical physics, which is based on operator algebras, as follows (but the same can be}} seen in Schr\"{o}dinger's picture):
From the KN viewpoint classical physics in terms of non-commuting objects:
$[q,p]=0$, but $[q, \pa/\pa q] =-1$ and
$[p, \pa/\pa p] =-1$,
where we use the operators that are needed to write classical equations of motion in the Hamiltonian (H) picture,
$\dot p = - \pa H/\pa q$ and $\dot q = \pa H/\pa p$.
{{In case of classical statistical physics, probability is emergent and Kolmogorov-like---additive, and without interferences.}}
Upon symmetrization of the commutators between these fundamental operators (repeating the procedure found in the context of spacetime physics) we get the fundamental commutation relations of quantum physics:
$[q,p]=i \hbar$ (where the central extension $\hbar$ is needed for dimensional reasons), and $[q, \pa/\pa q] =-1$ and
$[p, \pa/\pa p] =-1$.
{{Note that the probability measure is maximally symmetric, Euclidean (and called ``Fisher metric'' in statistics) and consistent with equations of motion. Thus quantum physics is endowed with an intrinsic probability that is not Komogorov-like. In particular, the compatibility of the Fisher metric for probabilities and the canonical equations of motion yields the quadratic Born rule, $|\psi|^2$, or equivalently, Gleason's theorem for density matrices, $\rho$, $\Tr(\rho)=1$, so that
$\langle O\rangle \equiv \Tr (\rho\,O)$, for any observable $O$, in general. 
Compatibly with the Born rule, only linear time evolution is allowed for the amplitudes of probability. This induces the fundamental interference of probabilities that is the hallmark of quantum theory, but {\it no} triple or higher order interference---owing to Born rule's (rigid) quadratic nature.}}

We can repeat the same logic for fields, such as in classical electrodynamics, by letting $q(t) \to \vec{A}(x)$, and  
obtain quantum electrodynamics, upon the same procedure of symmetrization of fundamental commutators.
The same procedure works for all local quantum field theories. Classical theories should be viewed via the correspondence principle as (in general, singular) limits of their deeper, quantum formulation.

Note: {{Spacetime physics}} (
gravity) is classical and 
has a {{fixed {\em polarization\/}}} (spacetime), but 
has {{dynamical geometry}} and is deterministic.
In turn, {{quantum physics}} (physics of matter) is intrinsically probabilistic and has a {{dynamical polarization}} $\psi \to U \psi$ ($U^\dag=U^{-1}$, which preserves $|\psi|^2$), but the {{geometry of quantum theory is rigidly fixed}}: It is the complex-projective geometry of projective Hilbert spaces, $CP^n$, the simplest being the Bloch sphere, $S^2=CP^1$. The Schr\"{o}dinger equation is thus the geodesic equation on $CP^n$, and the Born rule is idenfied with the Fubini--Study distance on $CP^n$.
{{By combining the two central featues of spacetime and quantum physics, we are led to suggest that a self-consistent quantum theory of {\em both\/} spacetime and matter (also known as quantum gravity) should have 
{\em both\/} a dynamical polarization 
{\em and\/} a dynamical geometry, and 
should be intrinsically probabilistic. 
That is, quantum probability in quantum gravity must be dynamical and, in general, contextual.}}

This in turn addresses the ``missing physics'' of the $c\hbar\GN$- and $\hbar\GN$-realms,
via
{{Quantum Gravity (QG) = Gravitized Quantum Theory (GQ)}}, with
{{a 
contextually dynamical
Born rule and quanta of spacetime. This view also fits into what is known to be true in classical general relativity.}}
Classical general relativity is {\em universal\/}: it ``gravitizes'' ({\em covariantizes\/}) all of classical physics, so it must not surprise that quantum gravity should ``gravitize'' quantum theory of quantum spacetime. Also, all previous theories, quantum mechanics, quantum field theory, as well as relativistic and non-relativistic views of spacetime physics, must be understood as limits of quantum gravity. This in turn will shed completely new light on the foundations of quantum mechanics, quantum field theory, as well as spacetime physics.

\section{Geometry of Quantum Theory and Beyond}
Here we present a
``bottom-up" reasoning for ``gravitization of quantum theory'' and elaborate on some points made in the 
introduction;
see~\cite{Minic:2002pd, Minic:2003en, Minic:2003nx}.  
As is well known, the
{geometry of quantum theory} (reviewed by Ashtekar and Schilling~\cite{Ashtekar:1997ud})
is described by maximally symmetric geometry of complex projective spaces, $CP^N$, with the Fubini--Study (FS) metric---which is the maximally symmetric Fisher metric of information theory~\cite{Wootters:1981ki}; see also~\cite{Braunstein:1994zz}. 
This metric reads
\be
\rd s_F^2 = \sum_i \frac{\rd p_i^2}{p_i}, \quad \sum_i p_i = 1,
\ee
where $p_i$ are probabilities. Let $p_i \equiv x_i^2$ so that we have a sphere in a Euclidean space
\be
\rd s_F^2 = \sum_i \rd x_i^2, \quad \sum_i x_i^2 = 1.
\ee
If $i=1,...,2(N{+}1)$, that is, if $x_i$'s are real and imaginary components of the wavefunction, then we have an odd-dimensional sphere $S^{2N{+}1}$, which is always a $U(1)$ bundle of $CP^N$ (with the FS metric). {The compatibility of the Fisher metric with the symplectic form underlying the equations of motion provides the complex structure---that is $i$ in the Schr\"{o}dinger equation~\cite{Minic:2004rj}.} It also demonstrates that probability is consistent with the equations of motion.
The overlap between the symplectic group responsible for the equations of motion and the orthogonal group responsible for the Fisher geometry of probability, gives the unitary group which governs the evolution quantum states and computes interfering quantum probabilities via the quadratic Born rule.

In this geometric formulation the Schr\"{o}dinger equation is the geodesic equation on $CP^n$ and
the Born rule is captured by the FS distance on $CP^n$:
\be
\rd s_{FS}^2 (1,2) = 4\big(1 - |\langle \psi_1| \psi_2 \rangle|^2\big).
\ee
Recall
that $CP^n = U(n+1)/U(n) \times U(1)$, where
$U(n+1)$ describes the unitary evolution (Schr\"odinger's equation),
$U(n)$ the non-Abelian Berry phase, and
$U(1)$ the complex phase of the wave function
; this captures the essentials of quantum mechanics.
Similarly, 
entanglement is represented by the Segre embedding of the product of lower-dimensional $CP^{n}$s into a higher-dimensional $CP^{N}$. 
({In the case of density matrices one has the Bures metric, an analog of the Fisher, or Fubini--Study metric}).
From the Schr\"{o}dinger equation,
and following Aharonov and Anandan%
~\cite{Anandan:1990Geo}, one can derive
\begin{equation}
    2 \hbar\,\rd s_{FS} = \Delta E\,\rd t,
\end{equation}
where $\Delta E$ is the dispersion of energy
and $\rd s_{FS}$ is the Fubini--Study metric on $CP^n$
(which is directly related to the Fisher metric). So, the robust, stable structure of quantum mechanics (including quantum field theory) is contained in the stability of the underlying Euclidean, Fisher metric of its statistical geometry. Note that the Schr\"odinger equation is just the geodesic equation on $CP^n$ sourced by the Hamiltonian ``charge'', which plays a kinematical role, as the generator of the unitary evolution. On the other hand, the Hamiltonian is also a gravitational source, and thus it has to play a dynamical role, suggesting a quantum version of the equivalence principle, and a dynamical nature of the quantum geometry. 

For fluctuating, quantum, spacetimes we expect topology change and hence, no unique timelike Killing vector, and thus $\Delta E$ becomes state dependent, so the geometry is state dependent, and therefore, dynamical~\cite{Minic:2002pd, Minic:2003en}. 
Also, at the Planck energy ($E_P$) scale $\Delta E = E_P$,
and thus $2 \hbar\,\rd s = E_P \,\rd t$, and given the dynamical time of general relativity, one expects a dynamical quantum metric $ds$~\cite{Minic:2004rj}.
{Therefore, we conclude}: Generally covariant quantum theory has a dynamical and contextual Born rule.
However, one might ask: why do we have to generalize the Euclidean Fisher metric to something dynamical, contextual and non-Euclidean?

The next important point is as follows: {Gravity violates assumptions of \v{C}encov theorem about the uniqueness of the Fisher metric from sufficient statistics~\cite{Berglund:2025qyo}. 
The
\v{C}encov theorem reads as: {\em Fisher metric is the unique information metric under sufficient statistics and identical independent measurements (i.i.d.).}
Here it is understood that:
({\small\bf1})~Data is a set of permanent results from independent identically distributed (i.i.d.) measurements---in quantum gravity the data is not i.i.d. but is instead non-Markovian, as each successive recording of a data point affects the next measurement: owing to the equivalence principle, storing data requires energy, which is a gravitational charge and thus back-reacts.
Similarly,
({\small\bf2})~sufficient statistics cannot be generated and passed between observers in quantum gravity due to different boundary conditions (gravity cannot be screened).

As a result of all these arguments, the issue of whether the Born rule in quantum gravity is fixed or dynamical, becomes an empirical question.

{\bf Experimental probes of dynamical Born rule and gravitized quantum theory:}
What would be the first experimental consequence of ``dynamical Born rule" or ``gravitized quantum theory''~\cite{Berglund:2025yoy}? 
The answer is as simple as it turns out to be elusive:
{intrinsic triple and higher-order quantum interference}! Such an experimental investigation of intrinsic triple interference in the context of gravity is indeed possible and feasible in the next few years, and such an experimental investigation would complement experimental attempts~\cite{Bose:2017nin, Marletto:2017kzi} that are searching for quantum entanglement of masses, because the measure of entanglement (``the entanglement witness'') in those experiments is also captured by the Born rule.
Note: {\em canonical\/} quantum theory (as routinely practiced) has {\it no} intrinsic triple quantum interference, owing to the rigid quadratic nature of the Born rule
and the associated fixed, maximally symmetric, geometry of the complex projective space. 

{In quantum mechanics and quantum field theory, any measurement of a triple interference, as a way of testing the Born rule, is a measurement of zero.}
{Current experimental bounds on the Born rule are all photonic and they are rather weak ($10^{-3}$). The reason for this weak bound lies, apart from experimental difficulties, in the lack of explicit proposals for the theoretical models of intrinsic triple quantum interference. That is why phenomenological work on modeling the variations of the Born rule is so important for the very foundations of quantum theory. Neutrino bounds expected to be
surprisingly similar (and are to be measured at JUNO)~\cite{Huber:2021xpx}.} 
{We emphasize that there are currently \underline{\em no} tests of the Born rule with gravity.}

Let us elaborate on this in more detail, following Sorkin~\cite{Sorkin:1994dt}: 
Classically, we have
addition of (independent) probabilities
\be
P_{n}(A,B,C,\cdots) \,=\, P_{1}(A) + P_{1}(B) + P_{1}(C) + \cdots\;,
\label{e:CMP}
\ee
for any number of paths. This reminds of the ``taxicab metric'' measurement of distance.
By contrast, the quantum-mechanical assessment of the probability for two paths is
$
P_{2}(A,B) = |\psi_A + \psi_B|^2 \vphantom{\Big|}$,
or more explicitly
\be
{|\psi_A|^2} + 
{|\psi_B|^2} + 
{(\psi_A^*\psi_B^{\phantom{*}} + \psi_B^*\psi_A^{\phantom{*}})}
\equiv P_{1}(A) + P_{1}(B) + I_{2}(A,B),
\label{e:QMP}
\ee
where the last term
\be
I_{2}(A,B) = P_{2}(A,B)-P_{1}(A)-P_{1}(B),
\ee
is the ``interference'' of the two paths $A$ and $B$.
{Thus, the non-vanishing double-path interference, 
$I_{2}(A,B)\neq 0$, distinguishes quantum theory from the classical one.
In contradistinction to~\eqref{e:CMP}, the quantum-mechanical probability computation~\eqref{e:QMP} reminds of the Euclidean distance, and is inherently and rigidly quadratic.
}

Precisely owing to this inherent quadratic nature, the {Born rule} dictates that all the superimposed paths interfere with each other---but only ever in a pairwise manner.
For instance, for three paths we have
$P_{3}(A,B,C) = |\psi_A {+} \psi_B {+} \psi_C|^2$, which evaluates to:
\be
P_{2}(A,B) {+} P_{2}(B,C) {+} P_{2}(C,A) 
{-} P_{1}(A) {-} P_{1}(B) {-} P_{1}(C),
\label{eq:three-slit}
\ee
where only pairwise interferences between the pairs $(A,B)$, $(B,C)$, and $(C,A)$ appear.
It is clear from the above that in order for there to be a non-linear correction in an interference pattern, the the rigid quadratic nature of the Born rule must be relaxed/generalized.  

Consider thus {\it a \underline{\smash{triple}} slit experiment}:
Since only pairwise interferences between the pairs $(A,B)$, $(B,C)$, and $(C,A)$ appear in~\eqref{eq:three-slit}, it makes sense to define any deviation from this relation as the intrinsic
triple-path interference: 
\begin{align}
\label{e:I3}
 I_{3}(A,B,C) \equiv P_{3}(A,B,C)
              &-P_{2}(A,B)
               -P_{2}(B,C)
               -P_{2}(C,A)\\
              &+P_{1}(A)
               +P_{1}(B)
               +P_{1}(C).\nn
\end{align}
(This straightforward to generalize for the case of $n$-paths.)
For both classical and quantum theory, this intrinsic triple-path interference is zero for any triplet of
paths. 
{Thus, experimental confirmation of $I_3=0$ is a confirmation of the (standard, rigid) Born rule.}  
This fact presents us with an exciting opportunity. One defines\cite{Huber:2021xpx}:
\be
\kappa = \dfrac{\varepsilon}{\delta}, \quad \varepsilon =  I_3(A,B,C), \quad \delta  =  |I_2(A,B)| + |I_2(B,C)| + |I_2(C,A)|.
\ee
Photonic experiments place only a weak bound ($\kappa \sim 10^{-3}$)\cite{Sinha:2010mwa, Park:2012jmy} (based on the experimental sensitivity, and not a particular model of intrinsic triple-interference), and no existing checks of the Born rule are known in gravity. Thus, such an experimental probe of the Born rule in the context of gravity is simply a must.

The claim~\cite{Berglund:2025qyo, Berglund:2025yoy}
is
that {with quantum gravitational degrees of
freedom turned on, one can get $I_3 \neq 0$,}
but for that one needs gravitized quantum theory, with
{observer dependent spacetime (Hilbert spaces) and dynamical, contextual, Born rule}.
The generalized probability in this approach to quantum gravity is given by analogy with non-linear optics, which, geometrically, 
resembles the Finsler metric,
\be
  P =    g_{ab}(\psi)\, \psi_a \psi_b \equiv \delta_{ab}\, \psi_a \psi_b + \gamma_{abc}\, \psi_a \psi_b \psi_c+\dots
   ,
\label{e:deformP}
\ee
where $a,b,c$ are state-space indices and with (schematically) a Schr\"{o}dinger plus Nambu quantum theory~\cite{Minic:2002pd, Minic:2003en}
\be
\frac{\rd \psi_a}{\rd \tau} = E_{ab} \psi_{b} + \Gamma_{abc}\, \psi_b \psi_c,
\label{e:deformDpsi}
\ee
where $\tau$ is the appropriate evolution parameter (and higher order generalizations).
As we have stressed, the Schr\"{o}dinger evolution is a geodesic equation on $CP^N$,
for a fixed (projective) Hilbert space, and so the above generalized equation should be a geodesic equation for a dynamical quantum geometry. This can be also understood as a proposal for a non-relativistic quantum $\hbar\GN$ gravity.

Here we include some comments on
Nambu mechanics which is based on volume preserving diffeomorphisms (also relevant for the dynamics of membranes and M-theory~\cite{Awata:1999dz}). The
Nambu bracket is a generalization of the Poisson bracket~\cite{Nambu:1973qe, Awata:1999dz},
\be
\{f, g, h\} \equiv \epsilon_{ijk}(\pa_i f)(\pa_j g)(\pa_k h).
\ee
The canonical example is a free asymmetric top whose equations of motion (Euler's equations for the rigid body) can be written as Nambu mechanics
\be
dO/dt \equiv \{H_1, H_2, O\},
\ee
where $H_1 = a L_1^2 + bL_2^2 + c L_3^2 \equiv E$
and $H_2 = L_1^2 +L_2^2 +L_3^2 \equiv L^2$.
The Nambu quantum theory is a generalization of the canonical Schr\"{o}dinger representation. Just replace $L_1, L_2, L_3$
by $\psi_1, \psi_2, \psi_3$, and 
the equation $\frac{\rd \psi_a}{\rd \tau} = \Gamma_{abc}\, \psi_b \psi_c$,
follows.
The matrix representation involves cubic and higher order matrices which nicely connects this subject to the subject of general probability theories (reviewed in, for example, in Ref.\cite{Plavala:2021wcu}).
Thus, deformations of the canonical Schr\"{o}dinger evolution by 
a Nambu-like quantum theory knows about $\psi_3$, the wavefunction that is responsible for
modeling  of ``quantum spacetime" in order to model
intrinsic triple interference. (Note that the above Schr\"odinger plus Nambu model is useful, even outside the context of quantum gravity, for placing the precise numerical bound on intrinsic triple interference of various experimental probes of the Born rule.)

Gravitized quantum theory in general expects
the complex projective state-spaces to acquire a considerably more complicated topology.
For example: the Bloch sphere may well become a Riemann surface of arbitrarily high (infinite) genus.
In general, this would imply higher order (cubic, etc.) terms in the expression for a dynamical Born rule which should involve new wavefunctions, such as $\psi_3$, ($\psi_1$ and $\psi_2$ being the real and imaginary parts of the canonical complex wave function $\psi$) etc.
Re-summing the infinite number of multilinear extensions could result in a euclidean GR-like (or, more generally, Finsler-like) theory
in the general space of states (that includes $\psi_3$, etc.); this is still under investigation.
Canonical quantum theory would be expected to emerge as a maximally symmetric limit, for example, by averaging over the infinite number of handles (handle-bodies) of the general quantum geometry.

{The important point here is that there {\em does exist} a non-linear optics analog of all this, where the effective triple interference {\em has been observed}~\cite{Namdar:2021czo}. In that case, the maximally symmetric Poincar\'e sphere of two polarizations in linear optics (which is the direct analogue of the Bloch sphere in quantum theory) gets deformed by including the third component of the electric field.}
Thus, effective triple interference is possible {\em and has been observed\/} in non-linear optical media.
(Instead of $\psi_1,\psi_2,\psi_3$, we have non-linear waves and the three components of the electric field $E_1, E_2, E_3$; instead of probability $P$, we have non-linear/cubic energy density $U$ that involves the product of all three components of the electric field, $U = \alpha_{ij} E_i E_j + \beta_{ijk} E_i E_j E_k$.)

{We claim that a similar ``smoking gun'' exists in the quantum case:}
{We claim that our Schr{\"o}dinger plus Nambu quantum theory deforms the well-known Talbot effect on a diffraction grating into its (multi-linearly) non-linear variant.}
{In this case the effect of decoherence is different from the observable signatures of the non-linear Talbot effect, which would involve the disappearance of certain fractional images of the diffraction grating. The experimental probes would be nano-particles, and the diffraction grating would be generated by a laser.}
{In view of the analogous experiment that was performed in non-linear optics, intrinsic triple interference with quantum gravity degrees of freedom should be analogous to a non-linear ``quantum spacetime medium''.}

{We also note that various previously attempted non-linear versions of quantum theory with fixed Hilbert spaces are \underline{\em not} the gravitized quantum theory we are describing, and thus they do tend to suffer from various problems. The crucial point is that generalized quantum geometry goes hand in hand with the quantum spacetime structure, and thus the usual problems associated with superluminal signaling simply do not appear.}
{Also, our approach indicates that in quantum gravity we have to go beyond traditional generalized probability theory given the non-Markovian nature of gravity due to its memory.}

At what scale should we expect to see triple-interference in our experimental setup?
As we will argue, the dual spacetime is relevant for the understanding of measurement in quantum theory. On the other hand, we will see that the geometry of the dual spacetime is the origin of dark energy. In what follows we will present the computation of the cosmological constant or the vacuum energy, the leading model of dark energy.
Our computation of the vacuum energy/cosmological constant (presented below), suggests a low energy scale of about $10^{-3}\,\text{eV}$, i.e., {$10^{-4}\,$m.} So that is the expected scale for the observation of intrinsic triple interference in the context of gravity. We now turn to the evaluation of the cosmological constant.

\section{Quantum Gravity and Observation: The Cosmological Constant (CC)}
\subsection{The CC Problem in QFT}
The cosmological constant (cc), $\Lambda$, (a parameter in Einstein's equations
$G_{\mu \nu} + \Lambda_{cc} g_{\mu \nu} = 8 \pi\GN T_{\mu \nu}$) has been measured using such different probes as supernovae, the cosmic microwave background (CMB) as well as the large scale structure.
It corresponds to the (quantum) vacuum energy $\Lambda_{cc}/(8 \pi G) \sim (10^{-3}\,\text{eV})^4$. 
The natural Planckian value, $(10^{19}\,\text{GeV})^4$, is $10^{124}$ times off---which encapsulates {\it the cosmological constant problem}.
So why is the universe so big, when it is, according to the current theory, supposed to be of a Planckian size? ({Conceptually, this is analogous to the problem of stability of atoms in classical physics, that opened the way to quantum physics.})
What is needed for the resolution of the cosmological constant problem is a theoretical approach that is flexible enough to explain the observed value of the cosmological constant and its
radiative stability.

In order to set up the cosmological constant problem in quantum field theory, let us start with the QFT vacuum partition function (free scalar):
\begin{equation}
    Z_{vac} = \int \rD[\phi]~ e^{-\int \frac{1}{2}\phi (-\pa^2 + m^2) \phi} 
    \propto \frac{1}{\sqrt{\det(-\pa^2 + m^2)}},
\end{equation}
which we can rewrite as
\begin{equation}
    Z_{vac} =  e^{-\frac{1}{2}{\rm Tr\, log}(-\pa^2 + m^2)}.
\end{equation}
In momentum space, $-\pa^2 = k^2$, and also
\begin{equation}
    -\frac{1}{2} \log(k^2+m^2) = \int \frac{\rd l}{2l} e^{-(k^2+m^2)l/2},
\end{equation}
where the Schwinger parameter $l$ is a worldline parameter associated with a {particle (quantum of the field $\phi$)}.  
Note that after taking the trace we have
\begin{equation}
    \int \frac{\rd^D k}{(2\pi)^D}\log(k^2 + m^2) = \int \frac{\rd^{D-1}k}{(2\pi)^{D-1}} \frac{\omega_k}{2},
\end{equation}
because
\begin{equation}
    \int \frac{\rd l}{2l}\int \frac{\rd k^0}{2\pi}e^{-(k^2+m^2)l/2} = \frac{\omega_k}{2},
\end{equation}
where $\omega_k^2 = k^2+m^2$ with $\omega_k$ equivalent to $k_0$ on-shell.
Thus, vacuum energy density in $D$ spacetime dimensions becomes
\begin{equation}
    \rho_0 = \int \frac{\rd^{D-1}k}{(2\pi)^{D-1}} \frac{\omega_k}{2} \sim \Lambda_D,
\end{equation}
where $\Lambda_D$ is the volume of energy-momentum space. ($\sum_k \frac{1}{2} \hbar \omega_k $, $\hbar=1$).
This is a divergent expression (to be regularized) that leads to {the cosmological constant problem}~\cite{Weinberg:2000yb}. 
The cosmological constant in 4d is, from Einstein's equations
$G_{ab} + \Lambda_{cc} g_{ab} =8 \pi G_N T_{ab}$, so that $\Lambda_{cc}\sim \rho_0 G_N \sim \rho_0 l_P^2$.

    Following the seminal calculation of Joe Polchinski note that the {vacuum partition function} is also 
    \begin{equation}
        Z_{vac} =\bra{0}e^{-iH\tau}\ket{0} = e^{-i\rho_0 V_D}
    \end{equation}
    where $V_D$ is the volume of $D$-dimensional spacetime, and $\rho_o$ is the vacuum energy density.
    Furthermore $Z_{vac}=\exp{(Z_{S^1})}$ where $Z_{S^1}$ is the partition function on $S^1$ in the world-line formulation
    \begin{equation}
        Z_{S^1} = V_D \int\frac{\rd^D k}{(2\pi)^D} \int \frac{\rd l}{2l}e^{-(k^2+m^2)\frac{l}{2}}.
    \end{equation}
    Thus the {vacuum energy density} is given by
    \begin{equation}
        \rho_0 = \frac{iZ_{S^1}}{V_d} \sim \Lambda_D,
    \end{equation}
and it is scaling with the volume of energy-momentum space as before. Here we witness a ``kinematical'' role of the vacuum energy (stemming from the fundamental generator of unitary transformations, the Hamiltonian $H$). However, the Hamiltonian and vacuum energy also source the gravitational field, and thus they should play a dynamical role as well, again suggesting a quantum version of the equivalence principle.

\subsection{The CC Problem in Quantum Gravity}
The cosmological constant problem persists even if we step outside the bounds of QFT into the domain of quantum gravity, here represented by string theory, which also naturally reduced to QFT at low energies. 
    As shown by Joe Polchinski, the computation of the cosmological constant proceeds as follows for the case of a {bosonic string}: Instead of one particle, we now have an infinite tower of particles with the following mass spectrum~\cite{Polchinski:1985zf, Polchinski:1998rq, Polchinski:1998rr}%
    ---the graviton ($h =1, \bar h =1$)
    \begin{equation}
        m^2 = \frac{2}{\alpha'}(h+\bar h - 2).
    \end{equation}
    Thus, summing over the physical string states
    \begin{equation}
        \sum_{\rm p.s} Z_{S^1} = \sum_{h,\Bar{h}} V_D \int \frac{\rd l (2\pi l)^{-D/2}}{2l}\int \frac{\rd\theta}{2\pi}e^{i(h - \bar h)\theta} e^{-\frac{2}{\alpha'}(h+\bar h -2)\frac{l}{2}},
    \end{equation}
    where we have imposed the level matching $h=\bar{h}$ (or $\delta_{h,\bar{h}}$).
If we define $\tau=\theta + i \frac{l}{\alpha'}\equiv \tau_1 + i \tau_2$, we get the partition function of a bosonic string on $T^2$
    \begin{equation}
        Z_{T^2} = V_D \int \frac{\rd\tau \rd\bar{\tau}}{2\tau_2} (4\pi^2 \alpha' \tau_2)^{-D/2} \sum_h q^{h-1} \bar{q}^{\bar{h}-1},
    \end{equation}
    where $q\equiv e^{2\pi i\tau}$. This can be derived directly from the Polyakov path integral. 
    Note that we can rewrite this expression, with $l\equiv\alpha'\tau_2$,
    \begin{equation}
(4\pi^2\alpha'\tau_2)^{-D/2} =\int\frac{\rd^Dk}{(2\pi)^D}e^{-k^2\frac{l}{2}}.
    \end{equation}
    Thus, as in QFT we can write  $Z_{T^2} \equiv V_D \int\frac{\rd^D k}{(2\pi)^D} f(k^2) \sim V_D \Lambda_D$ with 
    \begin{equation}
         \Lambda_D \equiv \int\frac{\rd^D k}{(2\pi)^D}; \quad f(k^2) \equiv \int_F \frac{\rd^2\tau}{2\tau_2} e^{-k^2 \alpha' \tau_2/2} \sum_h q^{h-1} \bar{q}^{h-1},
    \end{equation}
where $F$ is the fundamental domain.  
Note that $f(k^2)$ is dimensionless, so it does not contribute to the scaling of $Z_{T^2}$ and 
the vacuum energy $\rho_0 \sim Z_{T^2}/V_D$. The only difference is that in QFT the region of integration is 
   \begin{equation}
        |\tau_1|<\frac{1}{2},\quad\tau_2>0.
   \end{equation}
    In string theory, because of modular invariance (that is needed for consistency) 
    \begin{equation}
        |\tau_1|<\frac{1}{2},\quad |\tau|>1.
    \end{equation}
    So, the cosmological constant is UV finite in string theory, 
    but still $\rho_0\sim Z_{T^2}/V_D \sim \Lambda_D$---{and the cosmological constant problem persists in string theory!} 
    Note that unbroken supersymmetry (SUSY) implies zero vacuum energy, and broken SUSY brings back the problem, $\rho_0\sim \Lambda_D$.
    The AdS/CFT correspondence, or gravity/gauge theory duality~\cite{Maldacena:1997re, Gubser:1998bc, Witten:1998qj}, essentially provides fine tuning by turning on the fluxes which allow for a large AdS spacetime with a small {\em negative\/} cosmological constant, but nothing conceptually new appears in the above computation, neither from the AdS nor from the CFT side. Similar generic problems persist with compactifications of string theory, if one were able to compute the string partition function in that case, instead of using effective field theory arguments, and fine tuning, due to the presence of fluxes or other stringy solitonic objects.

Note that the large, even though UV finite vacuum energy, stems from its purely kinematical, quantum, role. But, the vacuum energy should play a gravitational, and thus, dynamical role as well, associated with the whole of spacetime, and its IR physics. How is that crucial UV/IR aspect of what we could call the quantum equivalence principle, to be captured? That is what we intend to answer in the next section.

\section{Resolving the Cosmological Constant Problem and QG=GQ}
\subsection{Quantum Gravity, Quantum Spacetime and Observation}

Here we describe the fundamental resolution of the problem encountered above and we point out how this resolution leads to the concept of gravitized quantum theory, from a ``top down point of view''.
Following Refs.~\cite{Freidel:2022ryr, Freidel:2023ytq}
(see also~\cite{Berglund:2022qsb})
we return to $Z_{S^1}$, setting $m=0$ for convenience and with $p$ denoting the momentum, 
    \begin{equation}
        Z_{S^1} = V_D \int \frac{\rd\tau}{2\tau}\int\frac{\rd^Dp}{(2\pi)^D}~e^{-\frac{p^2 \tau}{2}}.
    \end{equation}
    We observe that the spacetime volume is given by $V_D = \int \rd^D q$. We therefore consider the {following phase space} expression
    \begin{equation}
     Z_{S^1} = \int\frac{\rd\tau}{2\tau}~ Z(\tau),\quad
        Z(\tau) = \int\frac{\rd^D q}{(2\pi)^D}\int \rd^D p~
        e^{-\frac{p^2\tau}{2}} \equiv {\Tr} e^{-\frac{p^2\tau}{2}},
    \end{equation}
    where ${\Tr}$ is in {\it phase space}.
In $D=4$ we then get 
\begin{equation}
    Z(\tau) = \prod_{i=1}^4 \frac{1}{2\pi}\int_{-\infty}^\infty \rd q_i \int_{-\infty}^\infty \rd p_i~ e^{-\frac{p_i^2\tau}{2}},
\end{equation}
or by discretizing phase space
\begin{equation}
    Z(\tau) = \bigg(\frac{\lambda\epsilon}{2\pi} \sum_{k,\tilde{k}\in {\mathbf{Z}}} \int_0^1\rd x \int_0^1\rd\tilde{x}~ e^{-\frac{(\tilde{x}+k)^2\epsilon^2\tau}{2}} \bigg)^4,
\end{equation}
where $p\to \epsilon \tilde{x}, q\to \lambda x$ 
with $\lambda\epsilon= \hbar$.
This is divergent, but we can restrict the sum to finite range, {using modular regularization}, by following our papers on modular polarization in quantum theory and modular spacetime%
~\cite{Freidel:2015uug, Freidel:2016pls}.  
In what follows we will elaborate on this concept of modular spacetime, point out how it naturally appears both in foundations of quantum theory, but also in a chiral, phase-space-like, intrinsically non-commutative and T-duality covariant formulation of string theory called the metastring, which makes the quantum geometry associated with modular spacetimes dynamical, and it thus provides the resolution of the cosmological constant problem in string theory, as well as provides a ``top down'' reason for gravitization of quantum theory in quantum gravity. Returning to our computation, we have
\begin{equation}
    Z(\tau) = \bigg(\frac{\lambda\epsilon}{2\pi} \sum_{k=0}^{N_q-1} \sum_{\tilde{k}=0}^{N_p-1} \int_0^1\rd x\,\rd\tilde{x}~ e^{-\frac{(k+\tilde{x})^2\epsilon^2\tau}{2}} \bigg)^4,
 \end{equation}
 where $N_q,N_p$ count the number of unit cells in the spacetime and momentum space dimensions, respectively.  
 Now, define 
 \begin{equation}
     l\equiv N_q \lambda,\quad {\rm and}\quad \Lambda \equiv N_p \epsilon,
 \end{equation}
 where $l^4\equiv V_4$ is the size (volume) of spacetime and $\Lambda^4$ is the size (volume) of energy-momentum space, and $N=(N_p N_q)^4\in \mathbf{Z}$.
Thus, 
\begin{equation}
    l^4 \Lambda^4 = N,\quad {\rm or} \quad \Lambda^4 = \frac{N}{l^4}.
\end{equation}
   But there is actually an {upper bound on $\rho_0\sim \Lambda^4\leq \frac{N}{l^4}$} in $D=4$ due to the fact that $\exp(-p^2\tau/2)\leq 1$.
   {Also: $N$ should be thought of as entropy! One can see this as follows: according to the properties of modular space, one can be in a phase cell or not. Thus there are $2^N$ possibilities, and the logarithm of this number of possibilities, which represents entropy, scales as $N$.}
    The same bound also holds in string theory  following our earlier calculation of the partition function of the bosonic string on $T^2$ in $D=4$. ({Below we will show that the same bound holds in QFT and cosmological phase transitions described by an effective potential.)} Hence 
 \begin{equation}
    \rho_0 \leq \frac{N}{l^4}.
\end{equation}

Given the interpretation of $N$ as entropy, and given the fact that we are counting (for a fixed size of the momentum cell) the number of spacetime boxes, we now consider the {\it Bekenstein--Hawking bound}~\cite{Bekenstein:1980jp, Bousso:2002ju} 
 in a spacetime with a cosmological horizon, i.e., assuming that the cosmological constant is positive and we have a de Sitter (dS) spacetime. ({This famous  bound is a feature of semiclassical gravity, and also of gravitational thermodynamics.})
In static coordinates, dS spacetime metric is well-known
\begin{equation}
    \rd s^2_{dS} = - \Big(1-\frac{r^2}{r_{CH}^2}\Big)  dt^2 
    + \frac{dr^2}{\big(1-\frac{r^2}{r_{CH}^2}\big)}+r^2\rd\omega^2_{S^2}
\end{equation}
where $l\equiv r_{CH}$, the cosmological horizon, is the size of the observed spacetime.
{By identifying the above quantum number $N$ with the gravitational entropy},  the {Bekenstein--Hawking bound  ($S_{\text{grav}} = l_P^{-2} Area$)} becomes
\begin{equation}
    N\leq \frac{l^2}{l_P^2}. \quad 
\end{equation}
({Here we might think of possible experimental consequence for black holes, such as gravitational wave ``echoes'', because this reasoning can be applied to black holes as well and because N is also enormous in that case.})
Let us combine the Bekenstein bound with the bound on $\rho_0$ ($\rho_0\leq N/l^4$)
\begin{equation}
    \rho_0\leq \frac{1}{l^2 l_P^2}.
\end{equation}
{We are thus led to the mixing of the UV ($l_P$) and the IR ($l$) scales}.
Note that this scaling is compatible with Einstein's equations, but the point is that this scaling has been obtained via a quantum computation.
This is an explicit and well-known feature of the cosmological constant problem: one one side it knows about the infrared size of spacetime, and on the other about the ultraviolet degrees of freedom that contribute to the vacuum energy.
    With the cosmological constant in $D=4$ dimensions, $\Lambda_{cc}=\rho_0 l_P^2$ we then get the bound (that is model independent)
    \begin{equation}
        \Lambda_{cc} \leq \frac{1}{l^2}.
    \end{equation}
    Thus, the natural energy scale, $\epsilon_{cc}$ associated with the vacuum energy density (and model independent) is 
    \begin{equation}
        \rho_0 = \epsilon_{cc}^4\sim \frac{1}{l^2 l_p^2}.
    \end{equation}
    (Thus, the corresponding natural length scale, $l_{cc}\simeq 1/\epsilon_{cc}$), and we obtain the following \underline{\smash{\em see-saw formula}}. We will see that this formula is also true in full QFT (after cosmological phase transitions, for example)
    \begin{equation}
        l_{cc}\simeq \sqrt{l\, l_p}.
    \end{equation}
    This formula can be argued from a different point of view, and it can be shown to be radiatively stable in an explicit one-loop computation as shown in an ongoing work by Freidel, Kowalski-Glikman, Leigh and Minic.

    We note that the integration over the Schwinger parameter can be absorbed in the renormalization of the Newton constant. Also:
    \begin{itemize}
        \item the seesaw formula connects to reality by setting $l\sim 10^{28}\,$m because we obtain that $l_{cc}\simeq 10^{-4}\,$m or $10^{-3}\,\text{eV}$ in agreement with observations. We will use this logic for other predictions involving masses of fundamental elementary particles
        \item the seesaw formula is technically natural with $\rho_0\to 0$ when $l\to\infty$, and $l$ is the IR scale
        \item the seesaw formula indicates radiative stability, since we have no UV dependence, but this can be demonstrated explicitly
        \item the CC is small because the universe is filled with stuff (that is, the large number of degrees of freedom (dof) $N \sim 10^{124}$). Further exponentiation of this entropy $N$ sheds light on the extreme fine tuning of the initial state of the Universe (one part in $10^{10^{124}}$). This suggests more general quantum measures as precisely provided by gravitized quantum theory.
        \item $N$ is large because fluctuations scale as $\frac{1}{\sqrt{N}}$, and this fact is tied to the question of stability of the size of the observed universe (this is also reminiscent of Schr\"{o}dinger's famous argument regarding the question ``Why are atoms small?'')
        \item $N_i$ (where $i$ is $t,x,y,z$) is $N^{1/4} \sim 10^{31}$ (This can be interpreted as an Avogadro-like number for spacetime atoms and it might be of interest in the ongoing searches for quantum structure of spacetime in gravitational interferometry, as in the recent work \cite{Verlinde:2019xfb}).
    \end{itemize}
Furthermore we emphasize the {contextuality} of the argument: the measurement of a quantum observable depends on which commuting set of observable are within the same measurement set of observable, i.e., quantum measurements depend on the {\em context\/}.
\begin{itemize}
    \item First, $\epsilon$ is \underline{\em not} a cut-off, as $\epsilon$ and $\lambda$ can be arbitrary, though have to satisfy $\lambda\epsilon =\hbar$.
    \item Second, $\epsilon^4$ is replaced by $N$, which is the new quantum (``quantum gravitational'') number, and the size of spacetime, $l=r_{CH}$.
    \item Third, $N$ is determined by the Bekenstein-Hawking  bound--$N$ is related to $l$ and $l_P$, which is where gravity enters via $G_N\sim l_P^{2}$.
\end{itemize}
We stress that effective field theory (EFT) does not ``see" phase space and $N$ (EFT lives in classical spacetime, OR momentum space) and it does not mix UV and IR. (However, there exists interesting recent work%
~\cite{Becker:2020mjl}, as well as~\cite{Ferrero:2025ugd, Branchina:2025hen}.)

\subsection{EFT and the Cosmological Constant}
Let us look at the effective QFT partition function defined as
\be
Z(J) \equiv e^{i W(J)} = \int D \phi~ e^{i [S(\phi) + J\phi]},
\ee
where $W(J)$ is the generating functional of vacuum correlation functions, and it represents a direct analogue of
the partition function for a particle on a circle, or a string on a torus.
Given $W(J)$, we can define its Legendre transform to obtain
$\Gamma(\phi)$, the effective action, as
\be
\Gamma(\phi) \equiv W(J) - \int\rd^4x~ J(x) \phi(x).
\ee
The leading term in the expansion of $\Gamma(\phi)$ is the
effective potential,
\be
\Gamma(\phi) \equiv \int\rd^4 x [-V_{\text{eff}}(\phi) +\dots],
\ee
the minimum of which defines the vacuum energy in QFT.
Then by introducing the Schwinger parametrization we can obtain that
\be
\Gamma (\phi) \sim \int \rd^4 x \int \frac{\rd^4 k}{(2 \pi)^4}
\int \frac{\rd r}{r}~ e^{-U(k^2, \phi)\,r/2},
\ee
where the exponent in the above integral is bounded by 1. This argument can be repeated for QFT at finite temperature.
{By applying modular regularization to the crucial phase space factor together with the Bekenstein bound, we obtain the already derived result for the model independent bound on the vacuum energy.}

\subsection{From CC to the Masses of Elementary Particles:} Here we extend the computation of the bound on the vacuum energy to various bounds on the masses of elementary particles.
The essence of the vacuum energy calculation is contained in the evaluation of the vacuum partition function
$
        Z_{vac} =\bra{0}e^{-iH\tau}\ket{0} = e^{-i\rho_0 V_D} $
    where $V_D$ is the volume of $D$-dimensional spacetime, and 
    $\rho_o$ is the vacuum energy density, and the relation of the vacuum partition function to the partition function of particles (p) on the circle and the bosonic string (s) on the torus: $Z_{vac}=\exp{(Z_{S^1})}\equiv \exp{(Z_{p})}$ and : $Z_{vac}=\exp{(Z_{T^2})} \equiv \exp{(Z_{s})}$. 
Thus one gets that $\rho_0 \sim Z_{p,s}/V_D \sim \Lambda^D$ ($V_D \sim l^D$), as well as 
$
        Z_{p,s} \sim V_D \int\frac{\rd^D k}{(2\pi)^D}... 
        \sim l^D \Lambda^D.
$
{Then modular regularization of the phase space volume implies that $l^D \Lambda^D =N$,} where the physical meaning of $N$ is that is the entropy of spacetime degrees of freedom, and thus, using
holography, $N \sim l^{D-2}/l_P^{D-2}$.
In $D=4$ we get that $\rho_0 \sim 1/(l^2 l_P^2)$ and thus
$m_{\Lambda} \sim \sqrt{M M_P}$. (Here, $M$ is the Hubble scale, and $M_P$ is the Planck scale.) 
{$M=10^{-34}$\,{eV}; $M_P=10^{19}$\,{GeV}}.
Note, $M$ and $M_P$ are {contextual} IR and UV scales. {The general message of this computation is that we have to work with a dynamical phase space (because of dynamical spacetime) and thus, dynamical quantum phase space. This implies a dynamical Born rule, and a gravitization of quantum theory.}

The reader might say: ``So what? You have converted one scale, the scale of vacuum energy, into another scale, the IR scale set by the Hubble scale.'' We are in agreement with the reader, but we also note that the above computation can be extended to other elementary particles. Let us repeat the same logic for some {\it effective action}, that leads to the masses of elementary particles, instead of the one loop partition function for particles or strings. First comes the Higgs boson, for which the IR scale in the above computation is the vacuum energy scale, resulting in the Higgs mass $m_H \sim \sqrt{m_{\Lambda} M_P}$. One can also show the radiative stability of this value (thus shedding light on the hierarchy problem), as done in the upcoming paper by Freidel, Kowalski-Glikman, Leigh and Minic.

Then we can relate the mass of the top quark, the bottom quark and the tau lepton to the Higgs mass, by appealing to the Standard Model criticality (the minimum of the Higgs potential being given by the electroweak scale and the Planck scale) and the canonical renormalization group (RG).
Then the masses of other quarks and leptons, except for the neutrinos, follow from the masses of heaviest quarks and the heaviest lepton and the new emerging scales found from the matching of spacetime and matter entropies.
We obtain the first new relevant emerging IR scale, the \underline{\smash{\em Bjorken--Zeldovich scale}} (BZ), $M_{BZ}$,(see~\cite{Bjorken:2013aa, rBJ-MM}), by equating spacetime and matter entropies:
$N \sim l^2/l_P^2$ to $l^3/l_{BZ}^3$,
so that $l_{BZ}^3 \sim l\,l_P^2$. ($M_{BZ}^3 \sim M M_P^2\sim (7\,\text{MeV})^3$), which is the scale of the lightest quarks! The UV scale will be represented by the masses of the heaviest quarks, tied to the above mass of the Higgs boson.
Note that for neutrinos, given their masslessness at the classical level (due to parity breaking) we get by the same argument that the relevant emerging scale is given by the following matching of entropies $l^2/l_P^2 \sim l^4/l_{\nu}^4 \to l_{\nu} \sim l_{\Lambda}$. Therefore, the characteristic scale for neutrinos is the vacuum energy scale!

{So, the relevant IR scale for quarks and charged leptons is $M_{BZ}$ and the relevant UV scale is the heaviest fermion scale, which by Standard Model criticality~\cite{Froggatt:1995rt} is related to the Higgs mass.}
Given the linear nature of the fermionic equations of motion, as compared to the Higgs boson, two expressions for the fermion mass are possible: 
$
m_f \sim M_{\text{IR}} \sqrt{\frac{M_{\text{UV}}}{M_{\text{IR}}}}
\sim \sqrt{M_{\text{IR}} M_{\text{UV}}}
$
or
$
m_f \sim M_{\text{IR}} \sqrt{\frac{M_{\text{IR}}}{M_{\text{UV}}}}.
$
Therefore we end with the following list of results for various bounds of fundamental parameters in the Standard Model of particle physics:

First we have the cosmological constant, $m_{\Lambda} \sim \sqrt{M M_P} 
\sim 10^{-3}\,\text{eV}$ and then the Higgs mass,
$m_H \sim \sqrt{m_{\Lambda} M_P}
\sim 125\,\text{GeV}$. Given $m_H$ we determine $m_t$, $m_b$ and $m_{\tau}$ from $m_H$ and the Renormalization Group (RG) using SM criticality. Then we introduce the Bjorken-Zedovich scale, 
$M_{BZ}^3 \sim M M_P^2\sim (7\,\text{MeV})^3$, and the analogous neutrino scale, $l^2/l_P^2 \sim l^4/l_{\nu}^4 \to l_{\nu} \sim l_{\Lambda}$. This allows us to obtain the mass of the charm quark
\be
m_c \sim \sqrt{M_{BZ}\, m_t} = M_{BZ} \sqrt{\frac{m_t}{M_{BZ}}}
\sim 1.10~(1.27)\,\text{GeV},
\ee
and the mass of the strange quark:
\be
m_s \sim \sqrt{M_{BZ}\, m_b} =
M_{BZ} \sqrt{\frac{m_b}{M_{BZ}}}
\sim 171~(93.4)\,\text{MeV},
\ee
and the mass of the up quark:
\be
m_u \sim M_{BZ}^2/m_c \sim
M_{BZ} \sqrt{\frac{M_{BZ}}{m_t}}
\sim 10^{-2}M_{BZ} \sim 10^{-1}~(2.16)\,\text{MeV},
\ee
and finally, the mass of the down quark:
\be
m_d \sim M_{BZ}^2/m_s \sim
M_{BZ} \sqrt{\frac{M_{BZ}}{m_b}}
\sim 10^{-1} M_{BZ} \sim 1~(4.67)\text{MeV}.
\ee
Similarly, given the mass of the tau, we obtain the mass of the muon:
\be
m_{\mu} \sim \sqrt{M_{BZ}\, m_{\tau}} =
M_{BZ} \sqrt{\frac{m_{\tau}}{M_{BZ}}}
\sim 112~(106)\text{MeV},
\ee
as well as the mass of the electron:
\be
m_e \sim \frac{M_{BZ}^2} {m_{\mu}} \sim
M_{BZ} \sqrt{\frac{M_{BZ}}{m_{\tau}}}
\sim 464~(511)\,\text{keV}.
\ee
{Likewise, given the famous Weinberg dimension five operator as the expression that determines the mass of the tau neutrino mass (with the Standard Model scale $M_{SM}$ between $10^{14}\text{GeV}$ and $10^{16}\,\text{GeV}$ as indicated by the pairwise crossing of the three couplings, QED, QCD and weak), we have predictions for} 
({other two neutrino masses, given the tau neutrino mass}):
\be
m_3 \sim m_H^2/M_{SM} \sim (10^{-2} - 10^{-1})\,\text{eV},
\ee
that is, the muon neutrino:
\be
m_2 \sim \sqrt{m_{\Lambda} m_3} \sim (10^{-2.5} -10^{-2})\,\text{eV},
\ee
and the electron neutrino:
\be
m_1 \sim \frac{{m_{\Lambda}}^2}{m_2} \sim (10^{-4})\,\text{eV}.
\ee

We could extend these formulae to the 
CKM matrix elements (the quark mixing matrix)
\be
|V_{cb}| \sim \frac{M_{BZ}}{\sqrt{m_b\,m_d}}
\sim \sqrt{\frac{M_{BZ}}{{m_b}}} \sqrt{\frac{M_{BZ}}{{m_d}}}
 \sim 0.050 \quad (0.041),~
 (\leadsto\theta_{23})
\ee
as well as
\be
|V_{td}| \sim \frac{M_{BZ}}{\sqrt{m_b\,m_s}}
\sim \sqrt{\frac{M_{BZ}}{{m_b}}} \sqrt{\frac{M_{BZ}}{{m_s}}}
\sim 0.011 \quad (0.008),~
(\leadsto\theta_{12})
\ee
and
\be
|V_{ub}| \sim \frac{M_{BZ}}{\sqrt{m_b\,m_b}}
\sim \sqrt{\frac{M_{BZ}}{{m_b}}} \sqrt{\frac{M_{BZ}}{{m_b}}}
 \sim 0.002 \quad (0.003),~
 (\leadsto\theta_{13}).
\ee
By replacing the Bjorken-Zedovich scale with the vacuum energy scale
($M_{BZ} \to m_{\Lambda}$), as indicated by our matching of entropies of the spacetime and matter (massive and massless) sectors we obtain the
the neutrino mixing (PMNS) matrix:
\be
|U_{\mu 3}| \sim \frac{m_{\Lambda}}{\sqrt{m_3 m_1}} 
\sim \sqrt{\frac{m_{\Lambda}}{{m_3}}} \sqrt{\frac{m_{\Lambda}}{{m_1}}}
\sim 0.50 \quad (0.63),
\ee
as well as
\be
|U_{\tau 1}| \sim \frac{m_{\Lambda}}{\sqrt{m_3 m_2}} 
\sim \sqrt{\frac{m_{\Lambda}}{{m_3}}} \sqrt{\frac{m_{\Lambda}}{{m_2}}}
\sim 0.13 \quad (0.26), 
\ee
and
\be
|U_{e3}| \sim \frac{m_{\Lambda}}{\sqrt{m_3 m_3}} 
\sim \sqrt{\frac{m_{\Lambda}}{{m_3}}} \sqrt{\frac{m_{\Lambda}}{{m_3}}}
\sim 0.06 \quad (0.14).\ee
We note the obvious similar structure between the CKM and PMNS matrices, even though the numerical values of corresponding matrix elements are completely different.

{We have 20 parameters written in terms of Hubble and Planck scales (and the Standard Model scale). Are these results just coincidental? Or, is this indicative of the validity of the idea regarding gravitized quantum theory?}

\section{Quantum Gravity (QG)=Gravitization of Quantum Theory (GQ)}

{Suppose one does not think that above formulae are purely coincidental, and that they indicate that quantum theory should be gravitized. How does one formulate a theory of quantum gravity as a gravitized quantum theory of quantum spacetime?} 
The central intuition is to repeat the lessons of discovery of general relativity when gravity was introduced into relativistic physics, and when the geometry of spacetime was generalized from fixed Minkowski geometry into a dynamical geometry of general relativity. {Therefore, we want to view quantum theory of matter in classical spacetime as a special theory of quantum relativity and quantum gravity, the quantum theory of matter (visible and invisible) and spacetime, as general quantum relativity.}
This was done in
\cite{Freidel:2014qna, Freidel:2013zga, Freidel:2015uug, Freidel:2015pka, Freidel:2016pls, Freidel:2017nhg, Freidel:2017wst, Freidel:2017xsi, Freidel:2019jor}.

{In order to understand the deep structure of quantum theory as a special theory of quantum relativity we use the concept of modular variables} of Aharonov and collaborators~\cite{Aharonov:2005uc}, but in the context of spacetime. So, suppose we define our physics on a lattice (as in the lattice formulation of quantum field theory (QFT)). The lattice is just a discretized version of spacetime background and it is classical while the physics formulated on the lattice is quantum.
It turns out that in quantum theory we need (as noticed by Zak~\cite{Aharonov:2005uc, Freidel:2015uug}) both a {lattice ($l$) and a dual lattice ($\tilde{l}$)}. To see this we proceed as follows:

{\bf Modular variables and modular spacetime:}
Instead of considering the standard commutation relations between the position and momentum operators, $[q, p] =i\hbar$,
take the generators of translations in {phase space} (following the pioneering work of Weyl, Wigner and others)
\begin{equation}
    \hat U_a = e^{\frac{i}{\hbar} \hat p a},\quad
     \hat V_{\frac{2\pi \hbar}{a}} = e^{\frac{i}{\hbar} \hat q \frac{2\pi \hbar}{a}},\quad  \implies
     [\hat U_a, \hat V_{\frac{2\pi \hbar}{a}}] =0.
\end{equation}
In terms of {\it modular variables} \`a la Aharonov and collaborators~\cite{Aharonov:2005uc},
\begin{equation}
    [\hat q]_a \equiv \hat q~\mod a,\quad
    [\hat p]_\frac{2\pi \hbar}{a} \equiv \hat p~\mod\frac{2\pi \hbar}{a},
    \quad \implies
    [[\hat q]_a, [\hat p]_\frac{2\pi \hbar}{a}]=0.
\end{equation}
{Note that modular variables are covariant} (thus there is modular energy commuting with modular time as well).
Now take fundamental length $\lambda$ and energy $\epsilon$ on a phase space lattice, so that $\lambda \epsilon \equiv \hbar$.
{Modular variables are non-local (but consistent with causality, thus giving the origin of the uncertainty principle in this formulation)}.
{Similarly, modular variables are contextual}: {for example, in a double slit experiment the parameters $\lambda$ and $\epsilon$ are {contextual} and depend on
the experimental setup}. 

{Next comes the central point of viewing quantum theory as a special quantum theory of relativity. As in the case of special relativity with stems from two seemingly contradictory axioms of relativity of physics in inertial frames and the constancy of the velocity of light, quantum theory, endowed with intrinsic, fundamental probabilities, stems from compatibility of quantum spacetime with fundamental length/time and Lorentz symmetry (responsible for conservation laws of energy, momentum, etc). In special relativity we have observer dependent space and time, as ``slices'' of a new concept of spacetime (modular spacetime). In quantum theory, we have observer dependent spacetimes with maximally symmetric probability distribution consistent with equations of motion.
These observer dependent spacetimes can be viewed as ``slices'' of a quantum, modular spacetime, endowed with the fundamental non-commutativity $[x, \tilde{x}]=i \lambda^2$, which introduces the fundamental length/time.}

{Modular variables illustrate explicit non-locality}:  Take $H = \frac{p^2}{2m} + V(q)$ and write the Heisenberg equation of motion for $e^{ipR/\hbar}$, or equivalently $[p]_R$, where
$R$ is contextuality parameter, such as the distance between two slits in the double slit interference experiment\cite{Aharonov:2005uc}, 
\begin{equation}
    \frac{d [p]_R}{d t} = ...\frac{V(q +R/2) - V(q-R/2)}{R}.
\end{equation}
{This equation also illustrates that quantum mechanics arises from consistency between non-locality (from contextual modular variables) plus causality (compatibility with Lorentz symmetry, which controls causality)}.
Now, we are ready to reformulate the foundations and origin of quantum mechanics (QM) using (covariant) modular variables via modular spacetime.
({The new message here is that quantum theory tells us something new about quantum spacetime.})

So, let us introduce {\it modular spacetime}. First: what is {modular space}~\cite{Freidel:2015uug, Freidel:2015pka, Freidel:2016pls}?
{Modular space is the space of all commuting subalgebras of the Heisenberg--Weyl, or Weyl--Heisenberg, algebra.}
Note that $[q, p] =i\hbar$ defines what we call the Heisenberg--Weyl algebra, whereas $[[q]_a, [p]_{2 \pi \hbar/a}]=0$, which is valid for modular variables, defines commuting subalgebra of the Weyl--Heisenberg algebra.
Here the crucial input is provided by
{\it Mackey's theorem}\cite{Freidel:2015uug}: {``The space of all commuting subalgebras of the Heisenberg--Weyl algebra is a self-dual phase space lattice lifted to the Heisenberg--Weyl algebra''.}

If we use covariant modular variables,   {in other words, modular spacetime} of $d$ spacetime dimensions, we have the following quantum geometry (that we call Born geometry~\cite{Freidel:2013zga}) associated with Mackey's theorem.
First, we have the phase space or symplectic group, $Sp(2d)$ embodied in the symplectic form
$\omega_{ab}$, together with the self-dual lattice structure ($l$ plus 
$\tilde{l}$)---{that is, the doubly-orthogonal group} $O(d,d)$ embodied in the metric $\eta_{ab}$ (the symmetric counterpart of the antisymmetric 
$\omega_{ab}$). 
Then, 
to {define the vacuum} on this self-dual lattice---{we need a doubly metric structure} $O(2, 2d-2)$, denoted $H_{ab}$.
{The triplet of $\omega, \eta, H$ define the new concept of Born geometry}
\cite{Freidel:2014qna, Freidel:2013zga, Freidel:2015uug, Freidel:2015pka, Freidel:2016pls, Freidel:2017nhg, Freidel:2017wst, Freidel:2017xsi, Freidel:2019jor}
{The triple intersection of $\omega, \eta, H$, for a $d= (d-1) +1$ dimensional spacetime, gives the Lorentz group, illustrating the claim that the compatibility of the fundamental length/time in modular spacetime (which includes spacetime ($x$) lattice and the dual spacetime ($\tilde{x}$) lattice, stemming from the fundamental Weyl-Heisenberg algebra $[x, \tilde{x}]=i \lambda^2$), with the
Lorentz symmetry, leads to quantum theory.}
Thus {quantum mechanics follows from non-locality (fundamental length/time) consistent with causality (Lorentz symmetry) which implies a maximally symmetric probabilistic theory of superposed observer dependent spacetimes. (In general: dynamical causality in spacetime is expected to lead to gravitized quantum theory of quantum, modular, spacetime with dynamical and contextual quantum probabilities. We will argue that such theory of quantum gravity is the metastring formulation of string theory.)}

Note that given the commutative nature of modular variables, one can be localized in a particular phase space cell, 
thus making local QFT possible, 
but one can't tell in which phase cell, because the number operators that count the phase space cells do not commute with modular variables (and this is the reason for the uncertainty principle, and the existence of local operators of quantum fields, which obey the causal structure of the underlying classical spacetime).

{The reader might ask: How can fundamental length/time be intuitively consistent with Lorentz symmetry?} 
This is possible precisely because of {relative (observer dependent) locality}~\cite{Amelino-Camelia:2011lvm, Amelino-Camelia:2011hjg}. 
So: Different observers see different spacetimes (slices of modular spacetime),
{and different spacetimes have to be in linear superposition, so that fundamental length/time is indeed consistent with Lorentz symmetry.}
(This statement is very similar to what happens with spin: the superposition of up and down spin gives the Bloch sphere which is consistent with rotation symmetry, even though spin is discrete.) 
Therefore quantum theory does stem from quantum spacetime that is compatible with continuous spacetime symmetries, and the intrinsic probabilistic character of quantum theory is due to this fact, and quantum probabilities, consistent with equations of motion, have to interfere and obey the Born rule, at least in the situations in which the spacetime geometry is flat. Once gravity is turned on, the quantum probabilities ``warp'' and that is why quantum gravity is gravitized quantum theory of quantum spacetime.

\subsection{Modular Polarization, Metaparticles, Modular Fields and Observation}
The generic quantum polarization consistent with the structure of modular spacetime is~\cite{Freidel:2016pls}
is the {modular polarization} defined via the Zak transform~\cite{Freidel:2015uug, Freidel:2016pls}. Given Schr\"{o}dinger's $\psi_n(x)$ we
define
\begin{equation}
\psi_{\lambda} (x, \tilde{x}) \equiv \sqrt{\lambda} \sum_n e^{-2\pi i n \tilde{x}} \psi_n (\lambda(n+x)),
\end{equation}
(where $x \equiv q/{\lambda}$, $\tilde{x} \equiv p/{\epsilon}$, so $[x, \tilde{x}]=i$, $\lambda \epsilon = \hbar$).
From the point of view of the generic modular polarization, the more familiar Schr\"{o}dinger's polarization is very singular.
Introduce $\X^{A}\equiv (x^a , \tilde{x}_a)^{T}$, so that $ [ \hat{\X}^a, \hat{\X}^b] = i \omega^{AB}$.
We can now write the translations operators in phase space in a covariant form $W_{\K} \equiv e^{2 \pi i \omega( \K, \X)}$,
where $\K$ stands for the pair $(\tilde{k},k)$ and  $\omega(\K,\K')=k \cdot \tilde{k}' - \tilde{k}\cdot k'$.
($W$ is a generalization of the {Aharonov--Bohm phase---a prototypical example of modular variables~\cite{Aharonov:2005uc}}.)

So far we have discussed covariant quantum phase space as an example of modular space, but we have indicated that behind the foundations of quantum theory lies modular spacetime. How do we see that the modular spacetime is generic in quantum theory? One way is by realizing Born geometry in the description of a quantum particle.
Consider a {metaparticle} (mp) propagating in a modular space defined by Born geometry---$\omega, \eta, H$. The metaparticle~\cite{Freidel:2018apz} world-line action
$S_{\text{mp}} = \int\rd\tau\, L_{mp}$ (where the well known canonical particle is obtained in the limit of $\mu \to 0$ and $\tilde{p} \to 0$)
\begin{equation}
L_{mp} = p_\mu\, \dot x^\mu +\tilde p^\mu\, \dot{\tilde x}_\mu +\lambda^2 p_\mu\, \dot{\tilde p}^\mu - 
\frac{N}{2}\left(p_\mu p^\mu +\tilde p_\mu \tilde p^\mu - m^2\right) +{\tilde{N}}\left(p_\mu \tilde p^\mu - \mu \right),
\end{equation}
with the symplectic form $\omega$ in the ``Berry-phase'' term $p_\mu\, \dot{\tilde p}^\mu$, and the doubly-orthogonal metric $\eta$ in the diffeomorphism constraint $p_\mu \tilde p^\mu = \mu$
and the double metric $H$ is in the Hamiltonian constraint $p_\mu p^\mu +\tilde p_\mu \tilde p^\mu = m^2$.
We emphasize the explicit appearance of
{dual spacetime} $\tilde{x}$, $[x, \tilde{x}] = i \lambda^2$, and {dual momentum space} $\tilde{p}$, $[p, \tilde{p}] = 0$.
(Also, $[x, p] = i\hbar = [\tilde{x}, \tilde{p}]$.) {So spacetime $x$ is quantized, and it is canonically conjugate to dual spacetime $\tilde{x}$.

The metaparticle can be understood also as follows: If one ``second quantizes" Schr\"{o}dinger's $\psi(x)$ one naturally ends up with a quantum field operator ${\hat{\phi}}(x)$.
Similarly, the second quantization of the modular $\psi_{\lambda} (x, \tilde{x})$ would lead to
a modular quantum field (or quantum metafiled) operator~\cite{Minic:2020oho} ${\hat{\phi}}_{\lambda} (x, \tilde{x})$
\be
{\hat{\phi}}(x) \to {\hat{\phi}}_{\lambda} (x, \tilde{x}),
\ee
with $[x, \tilde{x}] = i \lambda^2$, leading to a covariant non-commutative field theory~\cite{Freidel:2017nhg, Freidel:2017wst, Freidel:2017xsi}. 
{As in generic non-commutative field theories~\cite{Douglas:2001ba} we expect a double renormalization group (RG) with an explicit mixing of the ultraviolet (UV) and infrared (IR) physics~\cite{Grosse:2004yu}. Thus modular field theory transcends the effective field theory (EFT) paradigm, and, as we shell see, it is directly related to the metastring formulation of quantum gravity, which contains metaparticles, the quanta of modular fields, as zero modes. Modular fields are in some sense remnants of metastring field theory, the metastring being an intrinsically non-commutative, T-duality covariant, chiral and phase-space like formulation of string theory in modular spacetime. 

Note that modular fields are generic if one has quantum, or modular spacetime, with canonical and dual spacetime, with observer dependent classical spacetimes, where the dual spacetime is necessary for understanding the intrinsic description of measurement and the classical limit in quantum field theory. In the limit in which the physics of dual spacetime is neglected, one recovers local quantum field theory. For example, by fixing the polarization of spacetime, in the fundamental commutation relations $[x, \tilde{x}] = i \lambda^2$, one recovers the usual local quantum field theory, in which all observers see the same background spacetime.}
Modular fields are contextual.
The classical spacetime label $x$ of canonical QFT corresponds to the choice of (classical spacetime) polarization in modular (quantum) spacetime with 
a contextuality parameter $\lambda$.
{By averaging over dual spacetime $\tilde{x}$ one can obtain classical evolution from the unitary evolution in the spacetime basis. This provides a new view on the quantum measurement problem not only in quantum mechanics but in quantum field theory.}

As is well known quanta of canonical quantum fields $\phi(x)$ are particles (and their antiparticles).
Similarly, 
{quanta of modular quantum fields $\phi_{\lambda}(x, \tilde{x})$ are metaparticles.} This is true in both the relativistic and non-relativistic context. Thus in condensed matter physics we have quasi-metaparticles~\cite{Barnes:2021akh}, which might be useful in modeling the relevant collective excitations of complex and strongly correlated electron systems, such as normal states of high-temperature superconductors.
{So, the first prediction of modular spacetime approach to quantum theory is the existence of metaparticles~\cite{Freidel:2018apz}.} 
Note that if we turn on backgrounds $p \to p + \phi$ and $\tilde{p} \to \tilde{p} + \tilde{\phi}$.
Thus we have {dual fields} {$\tilde{\phi}(x)$} in the effective classical spacetime $x$ description (after integrating over
the dual spacetime $\tilde{x}$ in order to account for the quantum measurement problem).
In what follows we will argue that dual particles, correlated to visible particles, represent dark matter.
{Visible $\phi$
and invisible (dark matter) ${\tilde{\phi}}$ do not commute and thus $\tilde{\phi}$ describes fuzzy dark matter, as we already known that the fields $\phi$ describe particles.} In principle, every Standard Model particle has a dual fuzzy counterpart. However, some dual fields, like the dual Higgs boson, might be relevant for modeling other phenomena, such as the Starobinsky-like inflation.

We can easily write an effective description of 
{dual ``particles'' (dual fields) that correspond to fuzzy dark matter}. To leading order in the fundamental length, that is, the contextuality parameter $\lambda$
\begin{equation}
    S_{\text{eff}} = - \int \sqrt{{g(x)}{\tilde{g}(\tilde{x})}} [R(x) + \tilde{R}(\tilde{x})+ 
    L_m (A(x, \tilde{x})) + \tilde{L}_{dm}   ( \tilde{A}(x, \tilde{x})) ],
\end{equation}
where the $A$ fields denote the usual Standard Model fields, and the $\tilde{A}$ are their duals, as predicted by the general (modular) formulation of 
quantum theory that is sensitive to the minimal length.
Note that we need to integrate over the dual space coordinates $\tilde{x}$ to get an effective
description of {visible matter, $A(x)$,  and dark matter, $\tilde{A}(x)$}, in classical $x$ spacetime~\cite{Berglund:2020qcu, Berglund:2021hbo, Berglund:2022qsb}.
This indicates that integration over dual spacetime variables (in the limit in which the fundamental length can be neglected) leads to a classical description. Note also that quantum gravity picks the natural basis for measurement and classical description - the spacetime basis. We comment on these important points in the conclusion as well.

Similarly, in the pure gravity sector we have dual geometry and dual curvature which nicely describes dark energy in classical spacetime.
To leading order in $\lambda$ we have
\begin{equation}
    S_{\text{eff}} = - \int \sqrt{-g(x)} \sqrt{-\tilde{g}(\tilde{x})} [R(x) + \tilde{R}(\tilde{x})+...].
\end{equation}
In this leading limit, the $\tilde{x}$-integration in the first term 
defines the gravitational constant $G_N$, and in the second term produces a {positive cosmological constant constant $\Lambda$}~\cite{Berglund:2020qcu, Berglund:2021hbo, Berglund:2022qsb}. 
{In general one can contemplate a time-varying dark energy~\cite{Hur:2025lqc}%
}.
{We emphasize that, in principle, visible and dark matter degrees of freedom are correlated (via the minimal length $\lambda$) which might serve as the physical origin (arising from fuzzy dark matter and its correlation to visible matter) of the so-called Milgrom's scaling in galaxies, clusters, superclusters and the so-called fundamental acceleration} $a_0 \sim c H/(2 \pi)$,
(with $\Lambda \sim H^2$)~\cite{Edmonds:2024qsj}.
The reason for this is that fuzzy dark matter is in the non-relativistic description sensitive to the quantum pressure, which in turn, is directly sensitive to the vacuum energy. Note that in the existing literature this observed fundamental acceleration scale (as well as the so-called acceleration floor) is associated with modifications of gravity, whereas our claim is that such observed acceleration scale is evidence for the existence of modular spacetime, modular fields, as well as the correlation of dual fields of fuzzy dark matter to the observed Standard Model fields of visible matter.

\subsection{Metastring Theory: An Explicit Realization of QG=GQ}

We stress that there exists an explicit realization in terms of an intrinsically non-commutative and T-duality covariant, chiral and phase-space-like reformulation of the bosonic string, the ``{\it metastring},''~\cite{Freidel:2014qna, Freidel:2013zga, Freidel:2015uug, Freidel:2015pka}. The action for the metastring is written below. 
There also exists a non-perturbative proposal~\cite{Berglund:2020qcu, Berglund:2021hbo, Berglund:2022qsb} 
of the metastring that is 
matrix model-like and, apparently, time-asymmetric, (with the following replacement in the action below $\pa_{\sigma} \cdot \equiv [\hat{\X}, \cdot]$, where $\hat{\X}$ matrix comes from a modular world-sheet~\cite{Berglund:2022qsb}). The elements of the matrix represent the non-local distinguishable spacetime quanta (where the local indistinguishable matter degrees of freedom emerge as ``boundary degrees of freedom'' from a particular quantum spacetime geometry). The metastring lives in modular spacetime and it is endowed with the Born geometry of generic quantum theory. Thus string theory stems from the foundations of quantum theory and it makes the geometry of quantum theory dynamical (gravitized). The metastring can be undersood as a proposal for a fully relativistic $c\GN\hbar$ quantum theory of gravity, viewed as gravitized quantum theory. The action of the metastring reads as
\begin{equation}
S^{\text{ch}}_{\text{str}}=
\int
\rd\tau\rd\sigma~
    \Big[\pa_{\tau}{\X}^{a} \big(\eta_{ab}(\X)+\omega_{ab}(\X)\big)
    -\pa_\sigma\X^a H_{ab}(\X)\Big] \pa_\sigma\X^b, 
\end{equation}
where $\X^a\equiv (X^a/\lambda ,\tilde X_a/\lambda )^{T}$ are 
coordinates on phase-space like (doubled) target spacetime and $\eta, H,\omega$ are all dynamical. $x^a, \tilde x_a$ come from the left and right moving modes of the bosonic string,
\begin{equation}
    x^a\equiv x^a_L + x^a_R,\quad \Tilde{x}^a \equiv \Tilde{x}^a_L - \Tilde{x}^a_R.
\end{equation}
In the context of a flat metastring we have 
 constant $\eta_{ab}$,  $H_{ab}$
 and $\omega_{ab}$ (zero $\omega_{ab}$ which provides a connection to the notation found in double field theory~\cite{Hull:2009mi})
\be\label{etaH0} 
	\eta_{ab} = \left( \begin{array}{cc} 0 & \delta \\ \delta^{T}& 0  \end{array} \right),\quad
H_{ab} =  \left( \begin{array}{cc} h & 0 \\ 0 &  h^{-1}  \end{array} \right),
\quad \omega_{ab} = \left( \begin{array}{cc} 0 & \delta \\ -\delta^{T}& 0  \end{array} \right).
\ee
The standard Polyakov string action is obtained when setting $\omega_{ab}=0$ and integrating out the $\tilde x_a$ ($h_{ab}$(X). Einstein's equation for gravity emerge from the zero beta function for $h_{ab}$, but there are also equations for the symplectic and doubly orthogonal backgrounds. The Polaykov action only knows about the $h$ background and not the whole Born geometry 
\begin{equation}
    S_P=\int \rd\tau\rd\sigma \gamma^{\alpha\beta} \pa_\alpha X^a \pa_\beta X^b h_{ab}(X)  + \ldots
\end{equation}

The triplet $(\omega,\eta,H)$ define the Born geometry~\cite{Freidel:2013zga, Freidel:2014qna}
and the metastring  propagates in modular, not classical, spacetime. 
Recall: the {\it space of commuting subalgebras of the Heisenberg algebra}, $[\hat x,\hat{\tilde{x}}]=i\lambda^2$, 
{\it defines modular spacetime}~\cite{Freidel:2015uug, Freidel:2016pls}
The new feature in the metastring formulation of the bosonic string is intrinsic non-commutativity 
and so there is a new  Heisenberg algebra (vertex operators provide a representation of the Weyl--Heisenberg algebra and one encounters no cocycles in the permutation of vertex operators)
\begin{equation}
    [\X^a,\X^b]=  i l_s^2 \omega^{ab} \implies [X^a,\Tilde{X}^b]=i\delta^{ab} l_s^2.
\end{equation}
(Note that intuitively, $\delta q \sim G_N \delta p \to \delta q \sim G_N \delta{\tilde{p}} \sim G_N \frac{1}{\delta{\tilde{q}}} \to \delta q \delta{\tilde{q}} \sim G_N \to [q, \tilde{q}] \sim i l_P^2$),
as well as the standard commutators (with $[\Pi, \tilde{\Pi}]=0$)
\begin{equation}
    [\X^a,\IP_b]=  i \hbar \delta^a_b \implies [X^a,\Pi_b]=i\delta^a_b\hbar,\quad 
    [\Tilde{X}^a,\Tilde{\Pi}_b]=i\delta^a_b\hbar.
\end{equation}
If the Kalb--Ramond $B_{ab}$ (which comes from the doubly-orthogonal transformation of the symplectic backgrounds) is constant, but non-zero, dual coordinates do not commute! In general, one has non-associativity. This is interesting in view of the observation regarding the origin of the Standard Model (SM) made by Gunaydin and Gursey~\cite{Gunaydin:1973rs}, in which the unique octonionic geometry $F_4/SO(9)$ of octonionic quantum theory appears and the Standard Model group emerges from the compatibility of the quantum phases
represented by $SO(9)$ in this case with the Poincare symmetry in 4 spacetime dimensions. Thus the familiar group $E_8$ of the heterotic string (which emerges from the bosonic string on a 16 dimensional self-dual lattice), might have a different interpretation as an octo-octonionic geometry which upon reduction to the real octonionic geometry leads naturally to $F_4$. Thus, from this point of view the SM gauge group emerges from the generalized phase of the effective octonionic geometry implied by the consistency conditions of heterotic string theory. The bounds on the masses of elementary particles of the SM computed in the previous section still hold (because they were independent on the Hamiltonian), and thus, they would fit in this proposal for the origin of the SM in the metastring formulation of the heterotic string.
{We note that zero modes of the metastring are metaparticles we have encountered in the discussion of the foundations of quantum theory based on modular spacetime}~\cite{Freidel:2015pka, Freidel:2018apz}. 
Modular fields are naturally associated with metastring string fields, and the metaparticles look like little rigid strings. According to our previous discussion, each SM particle has a correlated ``dual'' particle leading naturally to a fuzzy dark matter approach~\cite{Berglund:2020qcu, Berglund:2021hbo}.

{\bf Observing modular spacetime:} Given perhaps the surprising appearance of modular spacetime the reader might ask: how could we ``see'' the physical effects of modular spacetime?
One way would be that instead of scattering particles we should try to entwine them.
The reason for this is that 
vertex operators $V_K$ (that describe plane waves associated with asymptotic particle states) have co-cycles in the Polyakov string if we
assume that $[x, \tilde{x}] =0$
\be
V_{\IP} V_{\IP'} = e^{i (p \tilde{p'} - \tilde{p} p')}V_{\IP'} V_{\IP}.
\ee
The cocycle factor $e^{i (p \tilde{p'} - \tilde{p} p')}$ indicates the fundamental non-commutativity
of $x$ and $\tilde{x}$.
Can such a proposed entwining of particles be measured?
Here we are really suggesting to measure the effects of the ``R-matrix'', in the sense of ``swapping of particles'', 
instead of
the S-matrix, in the sense of ``scattering of particles'', which might be hard, but perhaps not impossible.

Another way of seeing the effects of the metastring formulation of a dynamical modular spacetime and thus, gravitized quantum theory, would be through the well-known 2d reduction associated with the high temperature behavior of string theory, that to leading order in the string length (the non-commutativity parameter), is inherited by the metastring. Such 2d reduction of gravitational degrees of freedom is associated with all other approaches to string theory and it is linked to ``asymptotic silence'' of classical general relativity. In the context of higher energy physics, such 2d reduction could be associated with a surprising appearance of planar 4-jets at the LHC at an effective energy of a TeV (this is being examined in an ongoing discussion between Nina Ilic and collaborators, as well as Dejan Stojkovic and Djordje Minic.).

One of the main point of our discussion of the metastring is that the metastring has a {dynamical Born geometry} $\omega_{ab}(\X),\eta_{ab}(\X), H_{ab}(\X)$, 
but Born geometry is the geometry of the modular spacetime formulation
of quantum theory. 
Thus {by making Born geometry dynamical the metastring ``gravitizes quantum theory''} (that is, {the metastring naturally makes the geometry
of quantum theory dynamical}). 
Given that the metastring is a theory of quantum gravity and that its background is quantum, modular, spacetime, we arrive at the conclusion that {``quantum gravity = gravitized quantum theory of quantum, modular spacetime''.}  {As we have already pointed out, the ``smoking gun'' signal of gravitized quantum theory is the dynamical nature of the Born rule for quantum probabilities, which implies triple and higher order quantum interference. This would be also the ``smoking gun'' for the metastring.}
Obviously, this reasoning is ``top-down'' and it complements our ``bottom-up'' reasoning presented in the introduction and the first section of this paper.
As already stressed, general relativity gravitizes all of classical physics, as it demands that all physics be consistent with the principle of general covariance. We claim that this empirically verified statement can be extended to the quantum domain by the procedure of gravitization of quantum theory.

{In particular: consider particle interactions as $0+1$ quantum gravity.}
As is well-known, quantum field theory can be viewed as $0+1$ quantum gravity.
Let us consider $\phi^3$ theory. The classical equations of motion read as
\be
(\pa^2 + m^2) \phi + g \phi^2 =0
\ee
{Given the interpretation of quantum field theory as a $0+1$  quantum gravity, the field $\phi$ can be viewed as the wave-function of $0+1$ universes and}
the above equation as a non-linear Wheeler-DeWitt equation, where the
interaction vertex stands for {topology change} in $0+1$ dimensions.
From a classical background or target spacetime viewpoint, we observe the decay of a particle described by the scalar field $\phi$, and we compute the decay by using the Born rule or, equivalently, the S-matrix.
However, from our QG=GQ viewpoint on $0+1$ universes (particles), {we have a natural triple correlation, associated with triple interference of $0+1$ universes.}
{Same for the ``pants diagram'' for strings---it represents intrinsic triple quantum interference from the $1+1$ dimensional quantum gravity point of view.}

In general, as in the case of gravitons, {whose interactions involve vertices of an arbitrary order, because of the non-polynomial nature of Einstein's gravity, we will have $n$-tuple quantum correlations with $n = 3,4,5,\dots$}. Thus, classical spacetime hides higher order quantum correlations, and in some sense, classical spacetime should be seen to emerge from such infinite order quantum interferences of quanta of spacetime (which can be viewed as fully dimensionally reduced metastring from what is naively a $25+1$ dimensional classical spacetime, on a self-dual lattice, and then {\it extensified} to a $3+1$ dimensional spacetime~\cite{Freidel:2015pka, Freidel:2017xsi}).

In principle, metastring provides and explicit theory of quantum gravity that involves quantum, modular, spacetime, endowed with the Born geometry of quantum theory, but, in general, dynamical and contextual, and thus gravitized, and so, implying more general quantum measures, and intrinsically triple and higher order quantum interference phenomena. The general equation for quantum Born geometry can be obtained from the requirement of conformal invariance of the metastring (or from the matrix RG equations in its non-perturbative formulation). The geodesic equations on the general Born geometry could be represented as the Schr\"odinger-Nambu equations that lead to triple and higher order quantum interference phenomena. At the same time, metastring has the relevant structure to reproduce the SM degrees of freedom and their dual, ``dark'' counterparts, as well as naturally small vacuum energy. The apparent time asymmetry of its non-perturbative formulation might be a quantum gravity source of CP violation beyond the SM. Note that supersymmetry is only an emergent property of the metastring. Presumably, the usual compactification viewpoint that starts from the 10 dimensional supersting (or its strongly coupled M-theoretic formulation) that, at least naively, leads to the swampland/landscape of string vacua with undetermined measures on that space of vacua, should connect to an extensified and fully quantum metastring with general quantum measures, but that is yet to be shown in detail.

\section{Quantum Spacetime and Quantum Gravity Phenomenology}

{We have already discussed the main phenomenological impact of viewing quantum gravity as gravitized theory of quantum spacetime. Here we want to point out that there are implications of this viewpoint on the quantum gravity phenomenology} {In particular, we emphasize that the statistics of spacetime quanta is different from the well-known statistics of matter quanta. Matter is granular and cuttable}: it consists of fermions that
are held together through interactions that are mediated by bosons (and exemplified in the famous the spin-statistics theorem of local QFT). 

In order to extend this atomic picture to spacetime we note the fundamental difference between spacetime and matter:
{Spacetime is extended and non-cuttable---at least ``away`` from spacetime singularities}. 
We claim that {spacetime quanta obey infinite statistics}
({or quantum distinguishable, or quantum Boltzmann statistics), captured by the Cuntz free algebra
$a_i {a_j^{\dagger}} = \delta_{ij}$} and non-commutative probability theory applicable to matrix models~\cite{Gopakumar:1994iq}, and are {held together by higher order quantum correlations responsible for higher order quantum interference effects.}
The huge number that appears in our discussion of the computation of vacuum energy can be identified with the spacetime Avogadro number $N \sim 10^{31}$ (per spacetime dimension) and the
vacuum energy formula $\delta \sim \sqrt{l\,l_P}$,
{can be related to efforts to detect a gravitational analog Brownian motion as an indicator of spacetime granularity via gravitational interferometry}~\cite{Verlinde:2019xfb}.

As is well known, Brownian motion can be understood by from the well-known formulae for fluctuations of various canonical statistical distributions. In the case of Bose--Einstein, BE, ($\bar{n}_k = (e^{(\epsilon_k - \mu)/T} - 1)^{-1}$), Fermi--Dirac, FD, ($\bar{n}_k = (e^{(\epsilon_k - \mu)/T} + 1)^{-1}$), and the Boltzmann statistics ($\bar{n}_k = (e^{-(\epsilon_k - \mu)/T})$), we can use the following basic formula for computation of the fluctuation of the number of particles $\Delta n \equiv n - \bar{n}$ (as reviewed in Landau and Lifshitz, volume 5, devoted to equilibrium statistical physics)
    \be
\langle (\Delta n_k)^2 \rangle = T \frac{\pa \bar{n}_k}{\pa \mu}.
    \ee
    Thus, we have, for BE: 
    \be
    \langle (\Delta n_k)^2 \rangle = \bar{n}_k (1+\bar{n}_k),
    \ee
    for FD: 
    \be
    \langle (\Delta n_k)^2 \rangle = \bar{n}_k (1 -\bar{n}_k),
    \ee
    and for Boltzmann's statistics: 
    \be
    \langle (\Delta n_k)^2 \rangle = \bar{n}_k,
    \ee
    which is precisely what is needed for the Verlinde--Zurek relation~\cite{Verlinde:2019xfb} found in the discussions of the gravitational analog of Brownian motion, after multiplying the number of particles with the corresponding energy $n_k \to E$, $(\Delta E)^2 = \bar{E}$.
    
    In the case of radiation BE corresponds to the Planck distribution and Boltzmann's statistics to its Wien's limit. Note that Wien's distribution is a purely quantum limit. We will see that it corresponds to the system being in the vacuum. In the Wien limit there is no stimulated emission. Thus, we have only absorption and spontaneous emission and absorption dominates. 
    We can understand this by recalling Einstein's $A$ and $B$ coefficients.
    Following Einstein, let us assume the balancing of spontaneous emission and stimulated emission (responsible for lasing) and absorption. Let us assume discrete levels with the population distribution determined by the Boltzmann--Gibbs distribution. The relative frequency of being in a state $m$
    with energy $E_m$ is given by $p_m e^{-E_m/T}$, where $p_m$ is that corresponding statistical weight. Spontaneous emission from the state $m$ to $n$ is (this is really responsible for non-commutativity of the creation and annihilation operators)
$p_m e^{-E_m/T} A^n_m$. The stimulated emission gives $p_m e^{-E_m/T} B^n_m \rho$, where $\rho$ is the ambient density of energy.
Similarly for the absorption from $n$ to $m$: $p_n e^{-E_n/T} B^m_n \rho$. By balancing absorption and emission Einstein obtained:
\be
p_m e^{-E_m/T} [A^n_m + B^n_m \rho] = p_n e^{-E_n/T} B^m_n \rho,
\ee
with the equilibrium condition (imposed by demanding that
when $\rho$ goes to infinity $T$ goes to infinity):
\be
p_n B^m_n = p_m B^n_m.
\ee
Then one gets the Planck distribution (BE for $\mu =0$, the photonic gas) 
\be
\rho = \frac{A^n_m/B^n_m}{e^{Em-E_n}/T -1},
\ee
provided that we have Bohr's relation $E_m - E_n = \hbar \omega$
and $A^n_m/B^n_m \sim \omega^3$ (which can be derived by counting modes of the electromagnetic field).
It also follows that the average number $\bar{n}$
\be
\bar{n} = \frac{1}{e^{\hbar \omega/T} -1} \to 
\frac{\bar{n}}{\bar{n} +1} = e^{-\hbar \omega/T},
\ee
(which if one neglects the $n$ in $n+1$, by neglecting stimulated emission, gives the Wien distribution) and
from which one can deduce the commutation relation for $a$ and $a^{\dagger}$ operators in the usual fashion 
$a^{\dagger} |n\rangle = \sqrt{n+1} |n+1 \rangle$ as well as
$a |n \rangle = \sqrt{n} |n-1\rangle$. The ratio of absorption to the
emission probabilities is the above ratio $n/(n+1)$.

The main point is that in the limit of the purely quantum Wien distribution, by above derivation the stimulated emission goes to zero, and one is only left with spontaneous emission and absorption, which should be appropriate for spacetime quanta described by the quantum Boltzmann statistics which looks like the purely quantum Wien distribution. Also the annihilation and creation operators do not have a commutator (or anticommutator) relation. They are free.
The only relation is given by $a a^{\dagger} = 1$. The BE and the FD cases can be understood as imposition of relations between spacetime quanta. This provides a new view of matter, which is not strange from the point of view of relative locality, that is, observer dependent locality, encountered in the quantum spacetime foundations of quantum theory. In that discussion it was crucial to have observer dependent spacetimes. The relation of these observer spacetimes could be interpreted by what we consider as matter (described by local quantum fields). Usually one thinks of spacetime as being produced through a relation of matter probes. That is also possible from the point of view of statistics. Think of local fermionic degrees of freedom connected by an open string. The string would follow infinite statistics, because of its distinguishable nature (as opposed to the fermionic quanta).
Notice that the BE and FD algebras appear as zero norm limits of the q-deformed algebra $ a a^{\dagger} - q a^{\dagger} a = 1$ that gives infinite statistics in the $q=0$ limit (as observed by Rob Leigh in discussions with Laurent Freidel, Jerzy-Kowalski Glikman and Djordje Minic, in their unpublished work on infinite statistics and quantum gravity).
    
    In the Einstein fluctuation formula, there is only particle (or quanta) piece for Boltzmann (infinite or quantum distinguishable)statistics.
    There is no wave or the field piece (which is represented by $+\bar{n}_k)^2$ in BE (commutators), or by $-\bar{n}_k)^2$ in FD (anticommutators). This wave part is indicative of the purely classical field contribution, and it cannot be present in a purely quantum description.
    For example, gravitons, as spin 2 bosons, would have the BE distribution, with both quantum and field contributions, but gravitons are just quanta of the weak gravitational field, not of spacetime itself. The non-local spacetime quanta, on the other hand, have the purely distinguishable, quantum Boltzmann, or infinite, statistics, without the classical, field, piece.

    Note that the fluctuation formula for the BE quanta says that their number could be infinite. The FE expression is obviously limited by either zero or one, and thus the Pauli exclusion principle. This would lead to anticommutation relations for the annihilation and creation fermionic operators. The infinite statistics allows for an infinite number of quanta. Actually, infinite statistics corresponds to the purely vacuum contribution, and it can be understood as a planar limit of a matrix model, for which the appearance of infinite statistics is quite explicit.

    The above results for the fluctuations of the BE distribution and the A and B coefficients follow from a calculation of Jordan 
    from matrix quantum mechanics. (This somewhat unjustly neglected work has been recently discussed in~\cite{Duncan:2007Pas}.)
    What Jordan observed is that for free fields, one
    has the spectrum of the bosonic harmonic oscillator
    $E_k = n_k + \frac{1}{2}$ and thus to compute the
    fluctuations one has to square that and deal with the sum 
    \be
    \sum_k \sum _l  (n_k + \frac{1}{2}) (n_l + \frac{1}{2}).
    \ee
    Then upon normal ordering one gets
    exactly the above formula for fluctuation in the case of BE statistics (after subtracting the constant term due to normal ordering). One can see where the quadratic (field) piece $n_k^2$ comes from and where the particle contribution $n_k$ arises (from the cross product of the oscillator and the vacuum contribution).
    Note that for the fermionic case (not considered by Jordan in the last section of his famous paper with Born and Heisenberg on matrix mechanics, even though he did independently discover the FD statistics) one has (given the $-\frac{1}{2}$ fermionic vacuum contribution)
    \be
    -\sum_k \sum _l  (n_k - \frac{1}{2}) (n_l - \frac{1}{2}).
    \ee
    Once again we can see how $-n_k^2$ arises and how
    the $n_k$ comes about (after normal ordering).

    Note that the Boltzmann statistics has only the $n_k$
    contribution, which comes from the vacuum contribution multiplied by the contribution coming from excitations, but with the field part completely suppressed. So, in the case in which the system is in the vacuum one gets the Boltzmann result and a truncated (to particles or quanta) form of Einstein's fluctuations. This meshes nicely with our computation of vacuum energy in~\cite{Freidel:2022ryr}.
    This also meshes well with the planar limit (vacuum) of the matrix model and the relation of the matrix model, in the planar limit, and infinite statistics (as in Gross and Gopakumar's paper~\cite{Gopakumar:1994iq} and references therein).
 One can conjecture that a matrix model (a la metastring with modular spacetime, and a modular world-sheet) is behind these properties on a more fundamental level.
Then a UV self-consistent cutting-off of quantum gravity would be implemented, because, like string theory, the metastring has a natural UV cut-off (due to modular invariance)
as well as the natural IR cut-off, given the doubled, modular, spacetime background. Apart from being able to
cut itself off in the UV self-consistently, quantum gravity should also cut itself off in the IR self-consistently, and that is possible in the metastring matrix model formulation. 

The importance of these observations is that instead of a Gaussian distribution found in the simplest models of canonical Brownian motion, in the case of gravitational Brownian motion, one should find a Wigner distribution
(the non-commutative probability analog~\cite{Gopakumar:1994iq} of the Gaussian distribution)
and thus a distinguishing signal from other noise sources. This should be relevant for finding a clear experimental evidence of atoms of spacetime.

What would be some implications of the existence of spacetime atoms?
We have already seen that they are implicit in the foundations of quantum theory and quantum field theory, and that they appear in the computation of the vacuum energy, and the related computations of the masses of elementary particles.
Here we offer an argument very similar to the one we made for the vacuum energy, but now, for a bound state made out of atoms of spacetime that satisfy quantum distinguishable statistics, and make a bound state with an entropy that scales as a black hole entropy!
As with vacuum energy we perform a modular regularization of phase space and we have in four spacetime dimensions
$l^4 \Lambda^4 \sim N$, where $N$ is the entropy
coming from the log of $2^N$ states we can have,
$l$ is the IR size of spacetime, and $\Lambda$ is the
characteristic energy scale. (The counting of $2^N$ states, coming from having two alternatives for each covariant phase space box, to be in the box or not to be in the box, is implied by the incompressibility of phase space, or equivalently, from the matrix model point of view, the fact that matrix eigenvalues repel, and thus they behave as ``fermions'' in phase space.) Note that if we insist on
the validity of Einstein's equations for the bound
state in question (which will turn out to be a black hole from its entropic properties) we have that
the energy density $\Lambda^4$ (the source of Einstein's equations, either external or produced by the gravitational field itself) scales in the Einstein
gravitational equations as the product of the scaling of the Einstein tensor, which scales as $1/l^2$, and
the inverse of the Newton constant, thus $1/l_P^2$.
Thus $\Lambda^4 \sim 1/(l^2 l_P^2)$.
Therefore, the entropy of that bound state of spacetime atoms with distinguishable statistics, $N \sim l^4 \Lambda^4$ scales as $N\sim l^2/l_p^2$, which is the
area law for black holes, where $l$ is the size of the bound state.
(Similarly, de Sitter space with its cosmological horizon would be another bound state of spacetime atoms with quantum distinguishable statistics. This directly connects to our computation of vacuum energy.)

The prediction here would be that
the entropy of a macroscopic black hole is quantized, which should have implications for observation. (The quantum number $N$ is enormous for a typical black hole: it is of the order of $10^{80}$). A quantum system so excited is chaotic and perhaps observable in gravitational wave echoes~\cite{Wang:2019rcf}? Note that other approaches predict such quantization, but not this modular space picture, and definitely not the connection with infinite, or quantum distinguishable statistics of atoms of spacetime.

\section{Conclusion}

In this talk we have discussed the new view on quantum gravity (QG) as a gravitization of quantum theory (GQ) of quantum spacetime.
We have discussed a new view on the origins of quantum theory and quantum field theory based on the compatibility of quantum spacetime and Lorentz symmetry. In the case of gravity this view implies a
{gravitization of quantum theory with a dynamical geometry of quantum theory (dynamical and contextual Born geometry)}.
In this generalized quantum relativity (quantum mechanics and quantum field theory being examples of special quantum relativity),
in principle, one has not only double, but also triple
and higher order interference allowed. This would offer a``smoking gun experiment" for our proposal.

We have discussed how quantum theory, a new probability theory of interfering probabilities, also offers a specific view on quantum spacetime in the form of modular spacetime (and the Born geometry of quantum theory).
There is something very new in this picture---the dual spacetime which does not commute with classical spacetime and if offers a new view on the problem of measurement as well as the quantum to classical transition.
There are immediate consequences of modular spacetime:
Particles (for example, the fundamental particle of visible matter as described by the Standard Model) are singular limits
of metaparticles---correlated particles and dual particles, where dual particles (necessarily fuzzy, and thus, field-like, because of the fundamental noncommutativity of visible fields, which describe visible particles, and dual fields) can be associated with (fuzzy) dark matter excitations. In the case of the quantum field description of such excitations one has modular fields, which are formulated in modular spacetime, and which are endowed with a contextuality parameter (the fundamental length/time), with noncommuting
quantum fields and dual (dark) quantum fields.
This picture extends for the geometry of dual spacetime, and the observed dark energy appears to first order as curvature of dual 
spacetime.

Metaparticles turn out to be zero modes of the metastring, an intrinsically non-commutative and T-duality covariant formulation of string theory that is chiral and phase-space-like and that lives in a dynamical geometry of modular spacetime. Such a formulation of string theory is rooted 
in quantum foundations, and its full dynamical form provides for a fundamental model of gravitized quantum theory of quantum, modular, spacetime, which also includes matter (and dual matter) degrees of freedom, originating from the quantum geometry of the metastring (that is, generically, non-associative).

In summary, the most important phenomenological implications of the QG=GQ picture as well as future work include:
    \begin{itemize}\vspace*{-1mm}\itemsep=2pt
    \item Triple (and higher order) quantum interference as the ``smoking gun'' signature of quantum gravity as gravitized quantum theory of quantum spacetime. Metastring theory as an explicit example of such a gravitized quantum theory of quantum, modular, spacetime.
        \item Computation of the vacuum energy and the cosmological constant (CC)
        based on the modular spacetime structure and dynamical phase space as arguably the only existing observed quantum gravity phenomenon. By taking the exponent of the result found for the gravitational entropy $N$, the same computation can be extended to address perhaps the greatest fine tuning problem in physics associated with the initial state of the Universe. This also implies a radically new quantum measure (given by an exponent of the exponent of an effective action) for the computation of quantum probabilities in quantum gravity.
         \item Implications of the computation of the vacuum energy for the Higgs mass and Standard Model fermion masses and mixing angles. Predictions of the neutrino masses. Time assymetry of the non-perturbative matrix-like formulation of the metastring as the source of CP violation beyond the SM?
        \item Metaparticles (zero modes of the metastring) and
        dual particles and dual fields as models of fuzzy dark matter with implications for the observed fundamental acceleration in galaxies, clusters of galaxies and other structures.
        (Dual Higgs as a model of Starobinsky inflation? Similarly, the axion is ubiquitous in string theory, and the metastring, which could serve as the QCD axion, but with a dual partner more suited for a fuzzy dark matter excitation.)
        \item Dark energy (a cosmological constant plus a possible time variable part) as the curvature of the dual spacetime, with observable implications for time dependent vacuum energy.
        \item Gravitational wave ``echoes''---as probes black holes as bound states of distinguishable spacetime atoms?
        \item Probes of distinguishable spacetime atoms from gravitational interferometry of gravitational Brownian motion.
        \item Searching for possible LHC signals of a generic 2d reduction of spacetime in quantum gravity.
    \end{itemize}

{In conclusion, we have two more important points. Quantum theory teaches us about: 1)~quantum spacetime and 2)~a new kind of non-classical probability}. 

The first point is not widely appreciated and the concept of dual spacetime that is conjugate, in the sense of canonical commutation relations, to classical spacetime is not the usual textbook stuff. Nevertheless, this is an essential component of quantum theory and 
{quantum measurement should be understood by invoking the new concept of dual spacetime}}.
Dual spacetime labels can be viewed as covariant, non-local, and non-commutative hidden variables. However, by picking the spacetime polarization, the dual spacetime looks completely invisible. Still, the dual spacetime is important, because, one can show
(following the well-known computations regarding the rewriting of the classical equations of motion in the Schr\"odinger picture, as reviewed in~\cite{Hubsch:2023qus}) that upon averaging over dual variables (to leading order in the fundamental length) classical equations follow. Essentially, the integration over the dual spacetime amounts to the overlap between the unitary evolution and the orthogonal transformations associated with the maximally symmetric probability measure, and this results in the symplectic transformations associated with the classical equations of motion. {Therefore, we expect that there should exist phenomenological probes of this ``classicalization'' that can be distinguished from the observed effects of environmental decoherence. Also, this viewpoint implies that the emergent measurement basis is the spacetime basis, as provided by the general structure of quantum gravity as gravitized quantum theory of quantum spacetime.}

{Similarly, quantum probabilities should find their application outside the pure domain of microscopic physics. After all, they are probabilities, and as such, like classical probabilities they should have wide applications in science. For example, complex and adaptive systems~\cite{Hubsch:2023qus} might utilize quantum-like and generalized quantum-like probability measure for purposes of more efficient information processing}. In other words, one might find emergent, quantum-like, or mock quantum theory, in complex and adaptive systems under certain circumstances.
So: should classical Kolmogorov probability always be used in generic stochastic non-equilibrium complex adaptive systems? We do not think so! If not (due to violations of 
\v{C}encov's theorem in complex adaptive systems) then
one can show (at least in theory) that {{emergent, or analog, or ``mock'' quantum theory}} (and its dynamical, ``gravitized'' version) might present more realistic and optimal description of efficient information processing in complex adaptive systems~\cite{Hubsch:2023qus}. 
This might be the reason for the existence of Nelson's stochastic approach or de~Broglie--Bohm's approach to quantum foundations, and it should be explored in various experimental situations. %

We also note that all available interpretations of quantum theory (reviewed for example by the Stanford Encyclopedia of Philosophy, in 
the following entry: \url{https://plato.stanford.edu/entries/qt-issues/}\cite{S.E.Phil:2022Phi}) 
appear as aspects of this new view point on quantum physics, which also leads to gravitized quantum theory. For example, the original insights of Bohr about complementarity of the spacetime coordination and conservation laws are precisely realized in the compatibility of quantum spacetime and Lorentz symmetry. This is also true for the insights of Aharonov regarding modular variables. The many-world
interpretation (especially in its original relative state form) can be seen to emphasize observer dependent spacetime. The stochastic Nelson formulation actually has the dual spacetime in its set-up (without perhaps realizing this point) in the form of the conjugate Brownian motion, and the related de-Broglie-Bohm interpretation illustrates the crucial compatibility of quantum probabilities with the equations of motion (in the derivation of the so-called quantum potential). All these interpretations can be understood from a more complete understanding of quantum theory and its gravitization that we have reviewed in this talk.

{\bf Acknowledgments: }
Many thanks to P.~Berglund, N.~Bhatta, D.~Edmonds, J.~Erlich, L.~Freidel, A.~Geraci, P.~Huber, S.~Hur, N.~Ilic, V.~Jejjala, M.~Kavic, J.~Kowalski-Glikman, R.G.~Leigh, D.~Mattingly, H.~Minakata, R.~Pestes, D.~Stojkovic, T.~Takeuchi, C.~Tze as well
as K.~Nikolic and S.~Pajevic for insightful collaborations and illuminating discussions.
TH is
grateful to the Department of Physics, University of Maryland, and the Physics Department of the University of Novi Sad, Serbia, for recurring hospitality and resources. DM is
supported by the Julian Schwinger Foundation and the U.S. Department of Energy (under
contract DE-SC0020262) and he thanks Perimeter Institute in Waterloo, Ontario, Canada, for hospitality and support.
DM also thanks the Institute of Quantum Optics and Quantum Information in Vienna, Austria, the Institute for Fundamental Physics of the Universe in Trieste, Italy, the KEK Theory Center and RIKEN Center for Interdisciplinary Theoretical and Mathematical Sciences in Japan, the Institute of Physics, Belgrade, Serbia, Perimeter Institute in Waterloo, Ontario, Canada and the Miami Physics Conference in Fort Lauderdale, Florida, for hospitality and opportunity to present different versions of this talk. Finally, thanks to Prof.~Branko Dragovich for the invitation to present this work at his 80th birthday celebration at the end of May of 2025, in Belgrade, Serbia.

\end{document}